\documentclass[preprint2,numberedappendix]{emulateapj}

\usepackage[T1]{fontenc}
\usepackage{natbib}

\usepackage{bm,times,color}
\graphicspath{{./fig/}}
\usepackage{amsmath}

\usepackage{lineno}

\newcommand{\EQ}{\begin{equation}}
\newcommand{\EN}{\end{equation}}
\newcommand{\EQA}{\begin{eqnarray}}
\newcommand{\ENA}{\end{eqnarray}}

\newcommand{\Fig}[1]{Figure~\ref{#1}}

{}
{}
{}

{}
{}
{}
{}
{}
{}
{}
{}
{}
{}
{}
{}
{}
{}
{}
{}
{}
{}
{}
{}
{}

{}
{}
{}

{}

{}
{}

\newcommand{\nn}{\mbox{\boldmath $n$} {}}

\newcommand{\G}{\,{\rm G}}

\newcommand{\K}{\,{\rm K}}

\newcommand{\s}{\,{\rm s}}

\newcommand{\Mpc}{\,{\rm Mpc}}

\newcommand{\GeV}{\,{\rm GeV}}

\begin{document}

\title{On the measurement of handedness in {\em Fermi} Large Area Telescope data}

\author{
  Julia Asplund$^{1,2}$
}
\author{
  Gu{\dh}laugur J{\'o}hannesson$^{3,1}$\thanks{E-mail:gudlaugu@hi.is}
}
\author{
Axel Brandenburg$^{1,2,4,5}$\thanks{E-mail:brandenb@nordita.org}
}

\affil{
$^1$Nordita, KTH Royal Institute of Technology and Stockholm University, Roslagstullsbacken 23, SE-10691 Stockholm, Sweden\\
$^2$Department of Astronomy, AlbaNova University Center, Stockholm University, SE-10691 Stockholm, Sweden\\
$^3$Science Institute, University of Iceland, IS-107 Reykjavik, Iceland\\
$^4$JILA and Laboratory for Atmospheric and Space Physics, University of Colorado, Boulder, CO 80303, USA\\
$^5$McWilliams Center for Cosmology \& Department of Physics, Carnegie Mellon University, Pittsburgh, PA 15213, USA
}

\date{\today,~ $ $Revision: 1.76 $ $}

\begin{abstract}
  A handedness in the arrival directions of high-energy photons
    from outside our Galaxy can be related to the helicity of an
    intergalactic magnetic field.  Previous estimates by
    \citet{2014MNRAS.445L..41T} and \citet{2015MNRAS.450.3371C} showed a hint of a signal present in the
    photons observed by the {\em Fermi} Large Area Telescope (LAT).
    An update on the measurement of handedness in {\em Fermi}-LAT data is presented
  using more than 10 years of observations.
  Simulations are performed to study the uncertainty of the measurements, taking into
  account the structure of the exposure caused by the energy-dependent
  instrument response and its observing profile, as well as the background
  from the interstellar medium.  
  The simulations are required to accurately
  estimate the uncertainty and to show that
    previously the uncertainty was significantly underestimated.
  The apparent signal in the earlier analysis
  of {\em Fermi}-LAT data is rendered non-significant.
\end{abstract}


\section{Introduction}

Most of the macrophysical processes around us show no statistical
preference of one handedness over the other.
A good counterexample
are cyclones on a weather map that have a counterclockwise inward 
spiral in the
northern hemisphere and a clockwise one in the southern.
These opposite spirals
correspond to opposite handednesses, but on average,
the number of cyclones in the northern and southern hemispheres are
nearly equal, so even in this case, the total or net handedness averages to zero.
By contrast, at the microbiological level, for example, there is a
global preferred handedness for all life on Earth
with amino acids being
levorotatory and sugars dextrorotatory \citep{Rothery}.
Even one of the four fundamental forces in nature -- the weak force,
responsible for the $\beta$ decay -- shows a global preferred handedness.
It produces electrons whose spin is anti-parallel to the momentum
\citep{1956PhRv..104..254L,1957PhRv..106..386F}.
One then says the electrons are left-handed or have negative chirality,
which is the Greek word for handedness.

The examples above illustrate that handedness can manifest itself in
a number of different ways.
Mathematically, handedness can be related to the existence of a
pseudoscalar.
Unlike ordinary scalars, which preserve their sign under mirror
reflection, pseudoscalars do change their sign under mirror reflection.
Similarly, ordinary or polar vectors preserve their direction under
mirror reflection, while pseudo or axial vectors change their direction
under mirror reflection.
An example is the rotation of a car's axle, which looks reversed
in a mirror.
Likewise, the curl of a velocity vector, i.e., the vorticity, changes
direction, and therefore the dot product of velocity and vorticity also
changes its sign in a mirror and is therefore a pseudoscalar.
The dot product of the gravity vector on the Earth's surface
and its global angular velocity is also a pseudoscalar.
Another example is the magnetic helicity, i.e., the dot product between
the magnetic vector potential and its curl, the magnetic field.
It plays a particularly important role, because it is a conserved quantity
in electrically conducting media \citep{BF84}.

\begin{figure}
  \centering
  \includegraphics[width=\columnwidth]{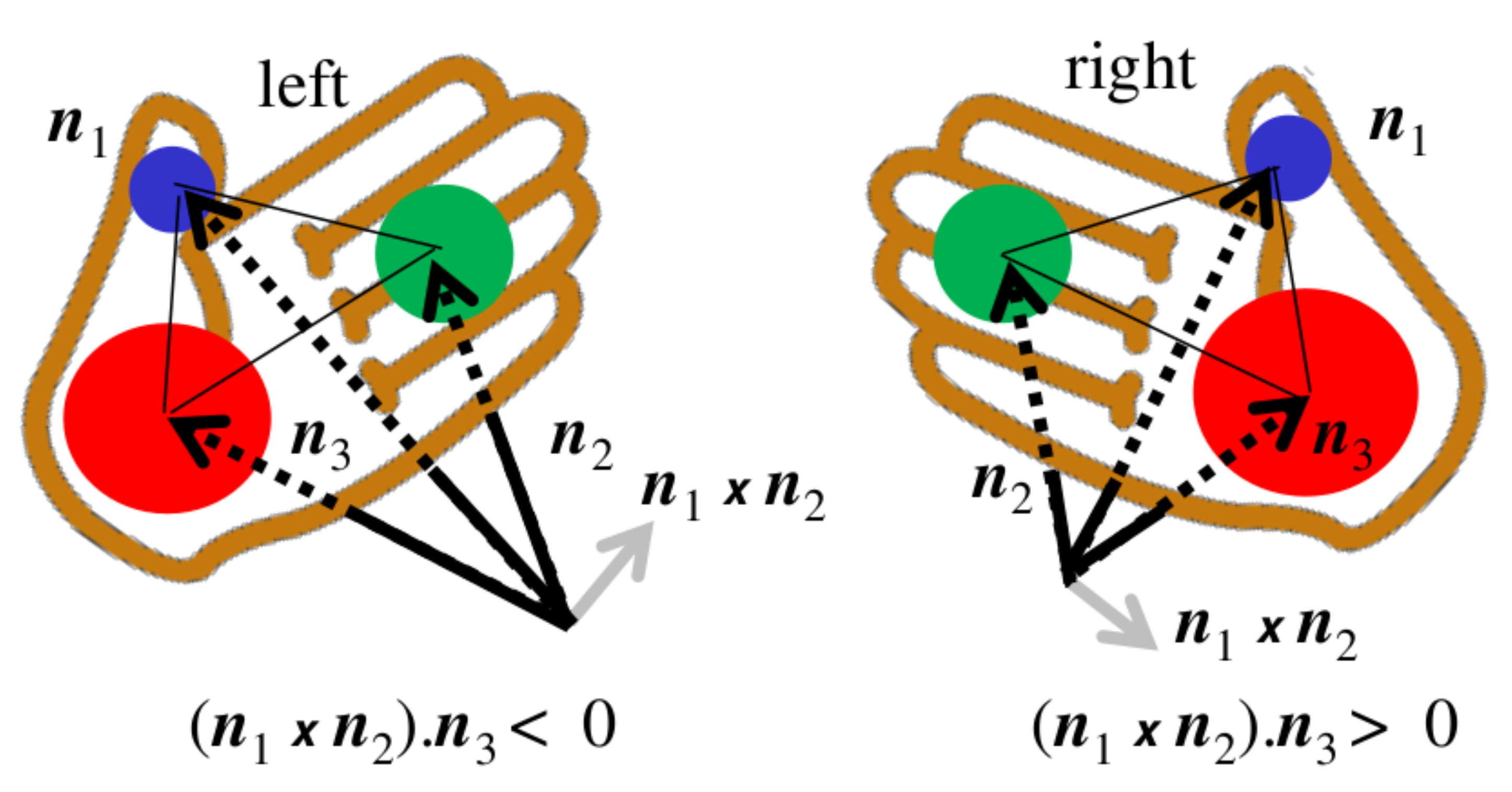}
  \caption{A sketch showing three patches of increasing size
  at positions $\nn_1$, $\nn_2$, and $\nn_3$ on a left and a
  right hand.
  The largest patch (red) corresponds to the palm of the open
  hand, the intermediate patch (green) corresponds to the fingers,
  and the smallest patch (blue) corresponds to the thumb.
  In this perspective view, the vectors $\nn_1$, $\nn_2$, and $\nn_3$
  start at the origins, which lie beneath each of the two hands toward
  their backsides.
  }\label{fig:Qsketch}
\end{figure}

Often, there is a causal connection between different pseudoscalars.
For example, gravity in a rotating body can cause finite kinetic and
magnetic helicities \citep{Mof78}.
Consider now the skew product
\EQ
Q=(\nn_1\times\nn_2)\cdot\nn_3
\EN
of three unit vectors
$\nn_1$, $\nn_2$, and $\nn_3$ of points on
a sphere; see \Fig{fig:Qsketch} for a sketch showing three patches
of increasing size (corresponding to larger energies)
at positions $\nn_1$, $\nn_2$, and $\nn_3$ on
a left and a right hand.
The largest patch corresponds to the palm of the open hand, the
intermediate patch corresponds to the fingers, and the smallest patch
corresponds to the thumb.\!\footnote{We thank the anonymous referee of
\cite{BB18} for suggesting this analogy in that paper.}
The two hands lie with their back on the sphere.
The cross product $\nn_1\times\nn_2$ of two polar vectors is an axial
vector, which points in the direction of $\nn_3$ for the right hand,
and in the opposite direction of $\nn_3$ for the left hand.
Therefore, $Q$ is positive (negative) for the arrangement of patches
on the right (left) hand.

A correspondence between the sign of $Q$ and the sign of magnetic helicity
was first proposed by \cite{2013PhRvD..87l3527T}.
They demonstrated the possibility of a causal link between the $Q$
product from the photon arrival directions on the celestial sphere and
the presence of a large-scale helical magnetic field permeating 
space even in the voids between galaxy clusters, far from any potential
astrophysical sources of magnetic fields.

The universality of the significance of this skew product was demonstrated
further by \cite{BB18}, who demonstrated numerically a connection between
the sign of magnetic helicity and the sign of the skew product for a
triple of magnetic spots on the surface of a sphere such as the Sun.
Thus, without necessarily relying on
a particular physical motivation
for the finiteness of the skew product $Q$ of 
unit vectors,
we wish to examine in the present paper the observational reality of a
possible detection.

To illustrate the far-reaching significance of a detection of net
handedness, let us mention here the connection between the possibility of
a globally helical magnetic field and baryogenesis in the early universe.
In fact, there is an epoch in the history of our universe during which
the weak force played a crucial role.
This is the time of the electroweak phase transition \citep{Vac91,V01}
some $10^{-11}\s$ after the Big Bang, when the temperature was
$10^{15}\K$, corresponding to an energy of $100\GeV$, or perhaps before
\citep{GB04}.
At that time, there could have been an excess of left-handed fermions.
Fermions and electromagnetic fields couple through the fine structure
constant in such a way that the total chirality of fermions and
electromagnetic fields does not change.
Moreover, the chirality of fermions destabilizes a weak magnetic field
and causes it to grow \citep{JS97,BFR12}.
This is in principle a promising mechanism for producing magnetic fields
of one sign of helicity throughout all of the universe.
However, already simple arguments \citep{BSRKBFRK17} suggested this would
only produce magnetic fields of $10^{-18}\G$ normalized to one megaparsec
length scale.
This might not suffice to explain the lower limit of magnetic field
of $10^{-16}\G\Mpc^{-1/2}$ \citep{NV10}, which is implied by the
non-detection of secondary photons from the halos of blazars.
On the other hand, doubts have been raised regarding the
exclusive need to explain this non-detection
by a magnetic field of a particular minimum strength
\citep{Broderick+18,Batista+19}.
In other words, significant levels of magnetic field may still exist, but
the minimum strength cannot reliably be constrained at the present time.

According to the theory of \cite{V01}, magnetic field
generation may have been accompanied by changes in the
Chern-Simon number to generate baryons, which would have implied
that the magnetic helicity also changes.
It has been shown, however, that the extraordinarily strong departures
from thermal equilibrium near the end of inflation \citep{GB99} could
have led to magnetic fields several orders of magnitude larger than what
can be estimated based on dimensional arguments \citep{GB08a,GB08b}.

If magnetic fields from the electroweak phase transition are to be
responsible for the lower limit of \cite{NV10}, they must have been
helical.
This is because only a helical magnetic field would have decayed
sufficiently slowly and would have increased its correlation length to
kiloparsec scales \citep{BKMRPTV17}.
The helicity of such magnetic fields can manifest itself in at least
three possible ways: in parity-odd polarization signals from the cosmic
microwave background \citep{KR05}, in primordial gravitational waves
\citep{KGR05}, and in the propagation properties of energetic photons
from blazars
that interact with the extragalactic background light
\citep{TVV12}.

Using a model of secondary particle emission from blazars,
\citet{2014MNRAS.445L..41T} developed a statistic
that could be applied to the
data from the {\em Fermi} Large Area Telescope (LAT).  Using the arrival
directions of photons from high Galactic latitudes in 60 months of LAT data,
they found an indication of left-handedness at the level of 3$\sigma$.
Interpreted in their framework, this indicated a left-handed magnetic helicity
for the cosmological magnetic field with a field strength of $\sim 10^{-14}$~G
on $\sim 10$~Mpc scales.  
In a follow-up analysis, \citet{2015MNRAS.450.3371C} also found
significant signals that persisted even when accounting for the effect of the
LAT energy-dependent exposure.
To test this claim, a new analysis is performed
using more than 10 years of LAT data with an updated event reconstruction
allowing for nearly a doubling of the statistics; see \cite{Asplund}
for a preliminary account of this work. In addition, simulations are
performed to check for the effects of the LAT energy-dependent exposure and
the contamination from the interstellar emission.  Using these simulations it
is found that the measured handedness is not significant.

\section{$Q$ Statistic and Data}

To look for a signal of handedness in the arrival directions
of GeV photons observed with the LAT, the $Q$ statistic
developed in \citet{2014MNRAS.445L..41T}
is used. 
The arrival direction of a photon measured to arrive at Galactic
longitude and latitude $(l,b)$ is represented by a unit vector in
Cartesian coordinates 
\begin{equation}
  \nn = (\cos b \cos l , \cos b \sin l , \sin b ).
  \label{eq:unitVector}
\end{equation}

The photons are binned in energy and the bins ordered
from low energy to high energy.  
Let $E_i$ denote the photons with observed
energies $E_{i,\min} < E < E_{i,\max}$ and say $E_i < E_j$ if $E_{i,\max} \le
E_{j,\min}$.  For any three energy bins, such that $E_1 < E_2 < E_3$, the $Q$
statistic is calculated as
\begin{equation}
  Q(E_1,E_2,E_3,R)=\frac{1}{N_3}\sum_{k=1}^{N_3}\left(\bm{\eta}_{1,k} \times
  \bm{\eta}_{2,k}\right) \cdot \nn_k(E_3),
  \label{eq:qstatistic}
\end{equation}
where $N_3$ is the number of photons in $E_3$, $\nn_k(E_i)$ is the unit
vector describing the arrival direction of photon $k$ in $E_i$, and
$\bm{\eta}_{i,k}$ is 
the mean of the unit vectors of photons 
in $E_i$ that
are within a radius of $R$ from the arrival direction of $\nn_k$, given
by
\begin{equation}
  \bm{\eta}_{i,k}=\frac{1}{N_i}\sum_{\nn_j(E_i)\in
  D(\nn_k,R)} \nn_j(E_i).
  \label{eq:eta}
\end{equation}
Here, $D(\nn_k,R)$ represents the circle of radius $R$ around
$\nn_k$ and $N_i$ the number of photons in $E_i$ that fall within the circle.
In case no photons fall within $D(\nn_k,R)$ and $N_i=0$, then $\bm{\eta}_{i,k}=0$.
Note that $\bm{\eta}_{i,k}$ is itself no longer a unit vector.
Using the mean vectors $\bm{\eta}_{i,k}$ significantly speeds
up the calculations compared to calculating the cross product
individually for each photon triplet. 
The ``standard error'' estimate for $Q$ is
\begin{equation}
    \delta_Q=\frac{\sigma_3}{\sqrt{N_3}},
    \label{eq:delta}
\end{equation}
where $\sigma_3$ is the standard deviation of the set used in the sum in
Equation~(\ref{eq:qstatistic}).  As will be shown later, this error estimate is
only appropriate in very limited situations.

To reduce the number of photons considered that originate
from the bright Galactic emission,
only photons observed near the Galactic poles are considered in the analysis.
From Figure~20 of \citet{2012ApJ...750....3A} it is clear that the
extragalactic background becomes dominant for $|b| > 60^\circ$
where the Galactic interstellar
emission becomes fairly constant with latitude.  Therefore, three different
latitude cuts are used to test the effect of the Galactic contamination, $|b| \ge 60^\circ$,
$|b| \ge 70^\circ$, and $|b| \ge 80^\circ$.  
This cut is only applied to the
$E_3$ photons, i.e., photons in $E_1$ and $E_2$ used in the analysis
can have an origin somewhat closer to the plane, $|b| \ge b_{\rm cut} - R$, where
$b_{\rm cut}$ is one of $60^\circ$, $70^\circ$, and $80^\circ$.  The latter
two values for the latitude cut are identical to those used in
\citet{2014MNRAS.445L..41T} while the $60^\circ$ cut is added to 
better characterize possible contamination and to increase the statistics.

The LAT is a pair conversion telescope, capable of observing photons in the
energy range from about 30~MeV to $>300\GeV$.
\citep{2009ApJ...697.1071A}.  Its wide field of view with half opening angle
of more than $60^\circ$ combined with a survey observing strategy makes its
$\gamma$-ray data set well suited for exploring handedness using the $Q$
statistic.  More than 10 years of P8R3 SOURCE class
\citep{2018arXiv181011394B} photon data from 1 Sept. 2008 to 1 April 2019 were
downloaded from the {\em Fermi} Science Support Center (FSSC)
\footnote{\url{https://fermi.gsfc.nasa.gov/cgi-bin/ssc/LAT/LATDataQuery.cgi}}. 
The P8R3 event selections are the most recently released data product from the LAT, providing
significant reduction in background compared to the previous P8R2 release.
Compared to the Pass 7 Reprocessed data used in \citet{2014MNRAS.445L..41T}, the
P8R3 also provides improved event reconstruction resulting in a narrower 
point-spread function and higher statistic.  For easy comparison with the results
of \citet{2014MNRAS.445L..41T}, the photons were binned in five energy bins from 10
to 60 GeV, each with a width of 10 GeV.  To simplify the discussion, the energy
bins will be referred to by their lower boundaries, e.g., `10 GeV photons' refers
to photons in the range 10--20~GeV.  The highest-energy bin will
always be the same, $E_3 = 50$~GeV, resulting in 6 combinations of energy bins
fulfilling $E_1<E_2<E_3$.  

Standard cuts were applied to the LAT data using
Fermitools version
1.0.1\footnote{\url{https://fermi.gsfc.nasa.gov/ssc/data/analysis/software/}},
including a maximum zenith angle cut of $105^\circ$ and a maximum rocking
angle of $52^\circ$ to reduce contamination from the very bright Earth limb
\citep{2019arXiv190210045T}.
Finally, events assumed to originate from
known point sources are removed. 
Many of them are extragalactic, but their emission does not originate in
  interactions with the extragalactic magnetic field, so their emission would 
reduce the signal-to-noise ratio for a helicity signal
\citep{2014MNRAS.445L..41T}. Therefore, a $2^\circ$ angular diameter region is
masked around every known point source given in the {\em Fermi}-LAT
Fourth Source Catalog (4FGL) \citep{2019arXiv190210045T}.  
A total of 585 sources in the 4FGL catalog are above $|b|=60^\circ$
resulting in about 33\% of the sky above $|b|=60^\circ$
being excluded by this cut.
Emission from the Sun and the Moon is non-significant for the 
energy ranges considered and is thus ignored.

After all cuts, the number of photons left to use in the analysis in each
energy bin above a latitude cut of $60^\circ$ are 
13740, 3478, 1558, 811, and 475, respectively as ordered in increasing energy. 
This is about a factor of 2 more in all energy bins
compared to the numbers presented in \cite{2014MNRAS.445L..41T} at the same latitude cut.
This is in agreement with the doubled observing time and larger acceptance of the
P8R3 dataset combined with the larger number of
sources in the 4FGL compared to the 1FHL (585 vs. 71 with $|b|>60^\circ$).  The reduced size of the exclusion
region around the point sources somewhat mitigates the solid angle lost to source cuts, but the excluded area
around the point sources in this analysis is still nearly 4 times larger.

\section{Synthetic data}

To test the accuracy of Equation~(\ref{eq:delta}), Monte Carlo simulations are
performed to estimate the statistical significance of the data.  
The uncertainty will be estimated as the standard deviation, $\sigma_Q$, of
the resulting distribution of $Q$ values that are calculated from each
simulated data set.
They are also
used to test for any possible bias in the handedness estimation caused by the
energy-dependent effective area of the
LAT\footnote{\url{https://www.slac.stanford.edu/exp/glast/groups/canda/lat\_Performance.htm}},
or the interstellar emission.  Three types of simulations are performed:
(i) arrival directions sampled uniformly on the sphere, (ii) isotropic
photon field accounting
for the LAT instrument response and observing profile, and (iii) interstellar
emission distribution of photon directions, also accounting for the instrument
response.  A combination of the latter two is used as the final error
estimate of the observations.  These simulations are described in the
following subsections.  
Emission from point sources is not included in the simulations, but the
  cut around the 4FGL sources is taken into account for the combined isotropic
and interstellar emission simulation for accurate error estimates.

\subsection{Uniform Photon Arrival Directions}

\begin{figure}
  \centering
  \includegraphics[width=\columnwidth]{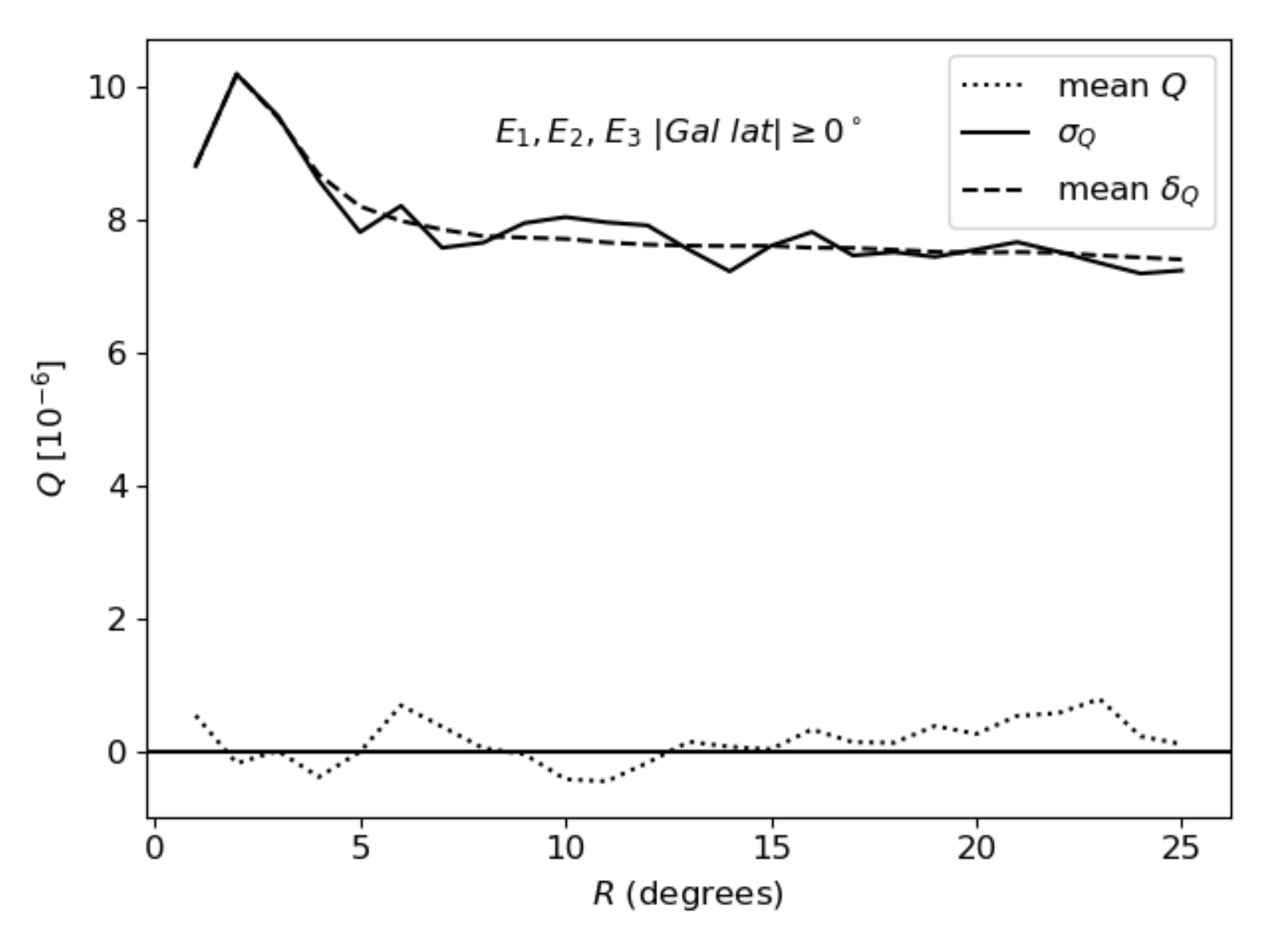}\\
  \includegraphics[width=\columnwidth]{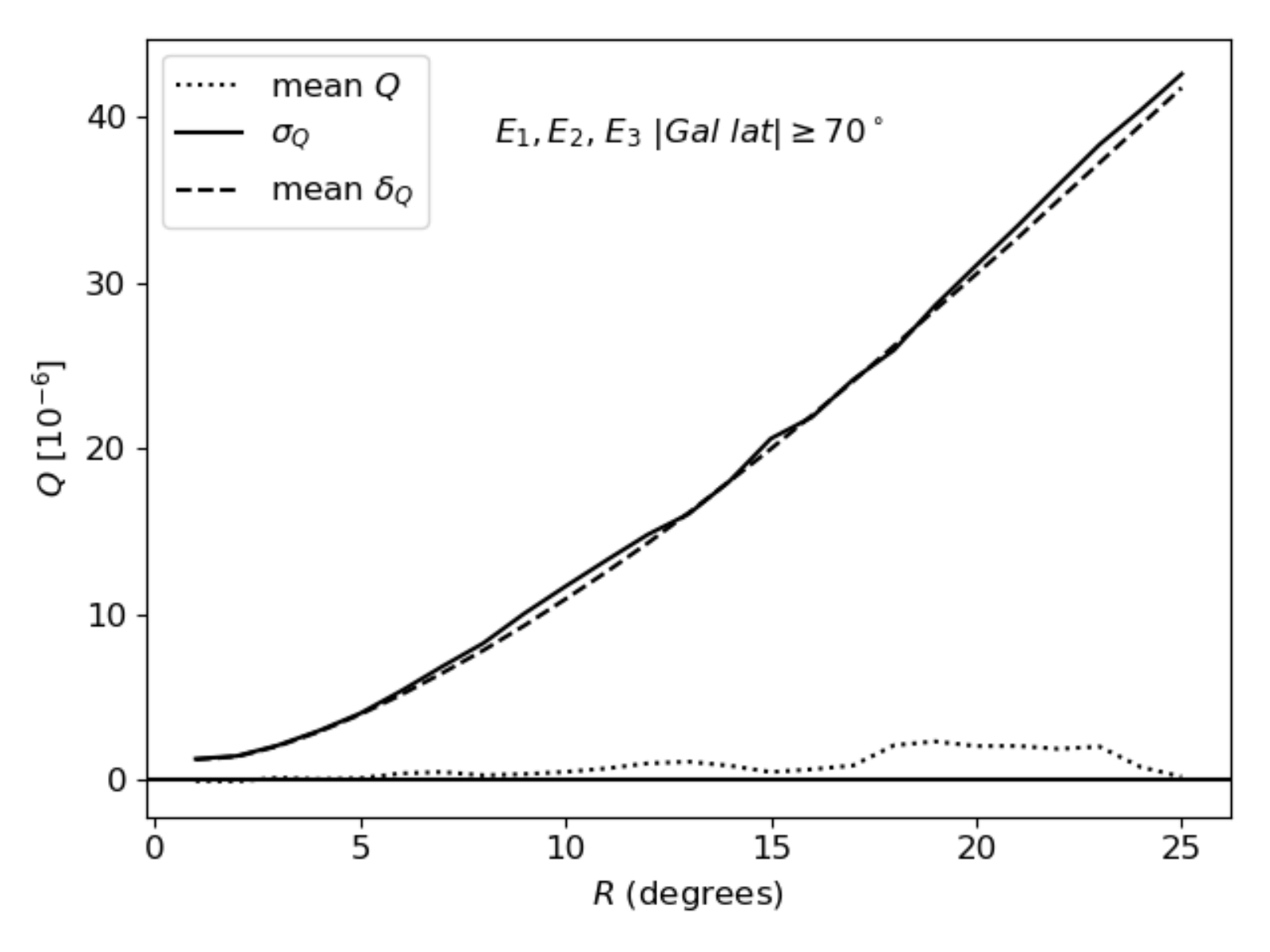}\\
  \includegraphics[width=\columnwidth]{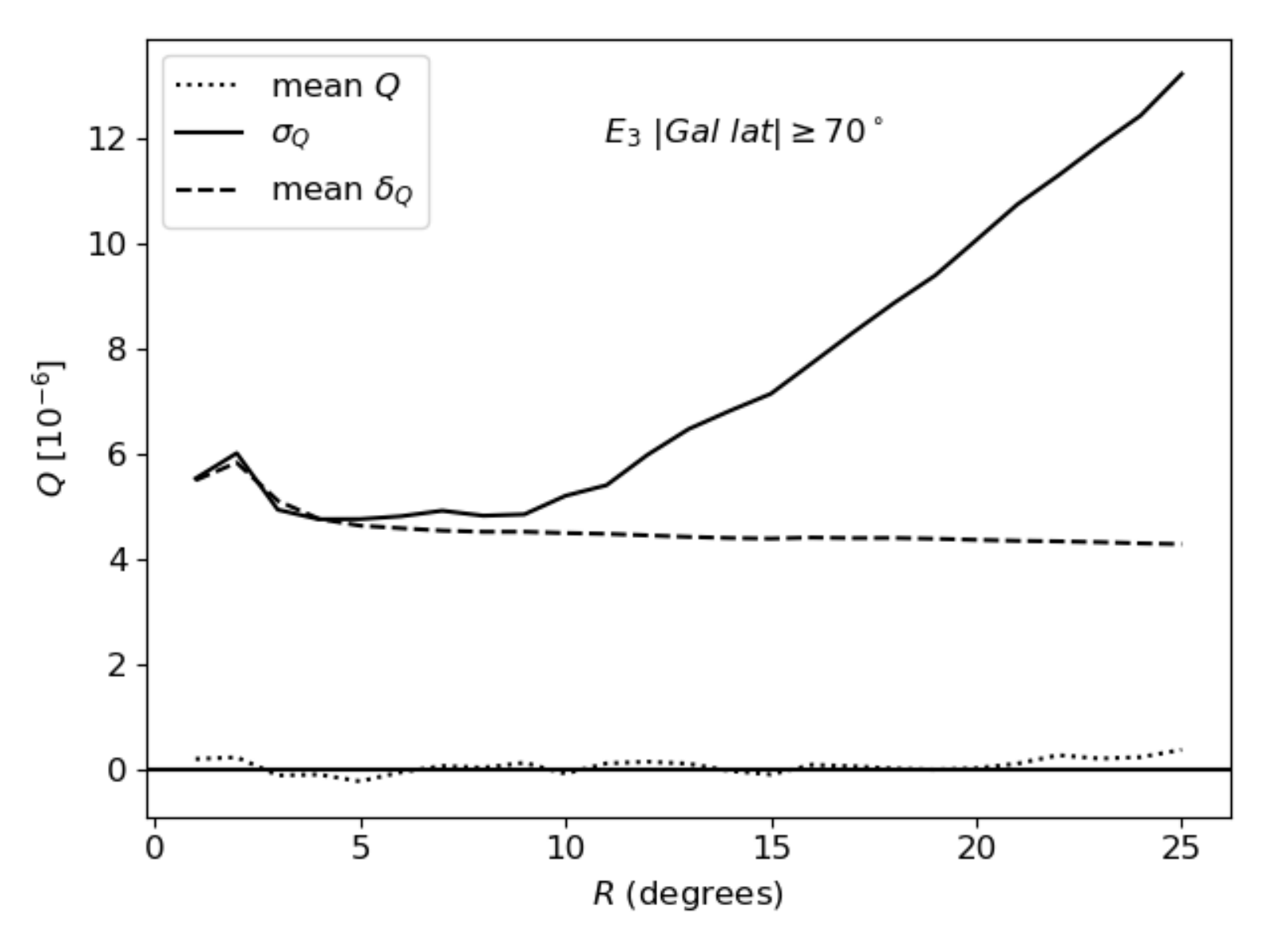}
  \caption{Summary statistics for the 500 Monte Carlo simulations using a
    uniform photon arrival direction using no latitude cut
  (top), a latitude cut of $70^\circ$ for all three sets (middle), and a latitude cut
of $70^\circ$ for $E_3$ only (bottom). Shown are for each $R$ the mean value of 
  $Q$ (dotted curve), the mean value of $\delta_Q$ (dashed curve), and the standard deviation $\sigma_Q$
(solid curve) of the obtained distribution from the simulations.  
$R$ is the radius of the circle used to select the low energy photons,
see the text for details.
  }
  \label{fig:uniformstats}
\end{figure}

\begin{figure*}
  \centering
  \includegraphics[width=0.49\textwidth]{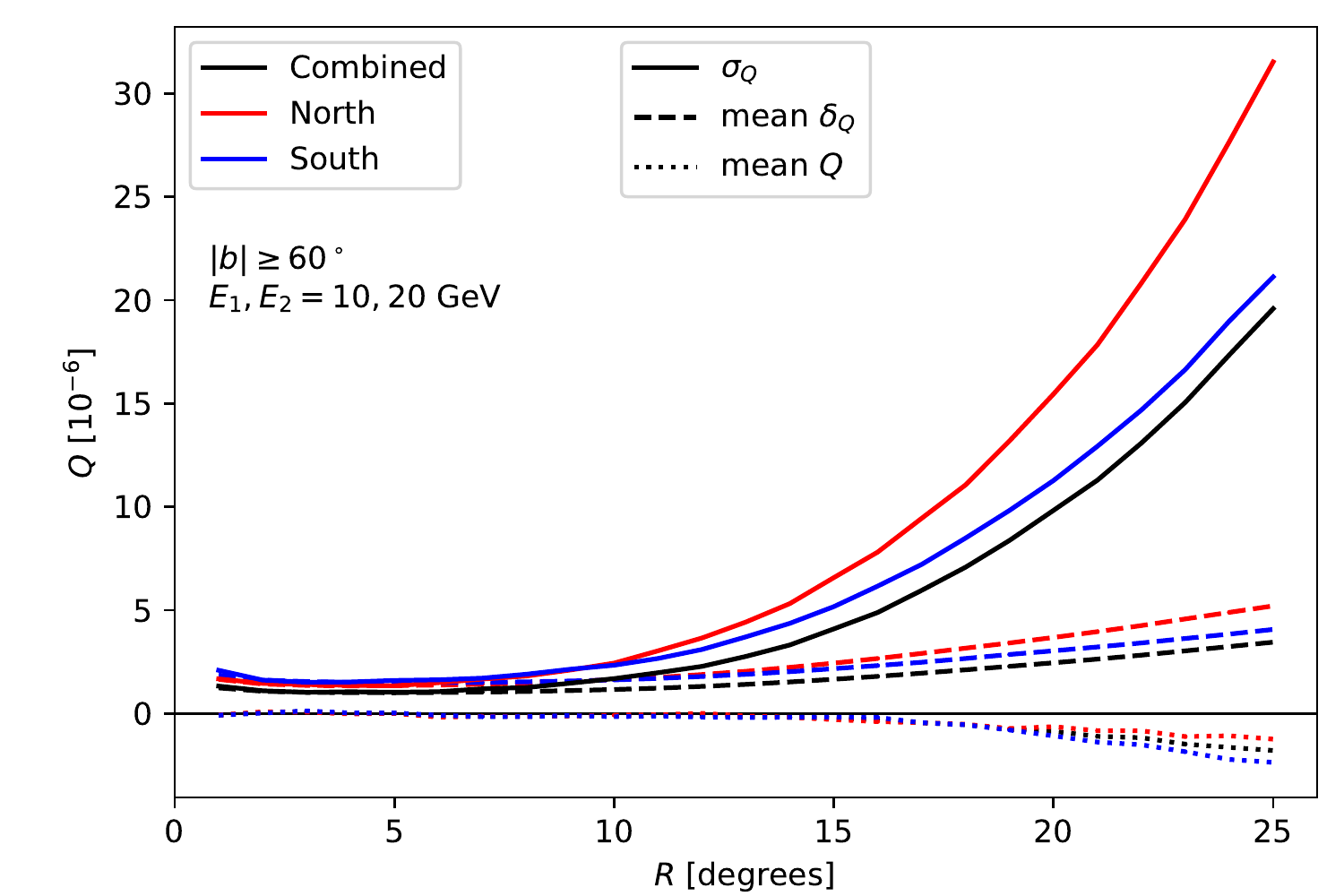}
  \includegraphics[width=0.49\textwidth]{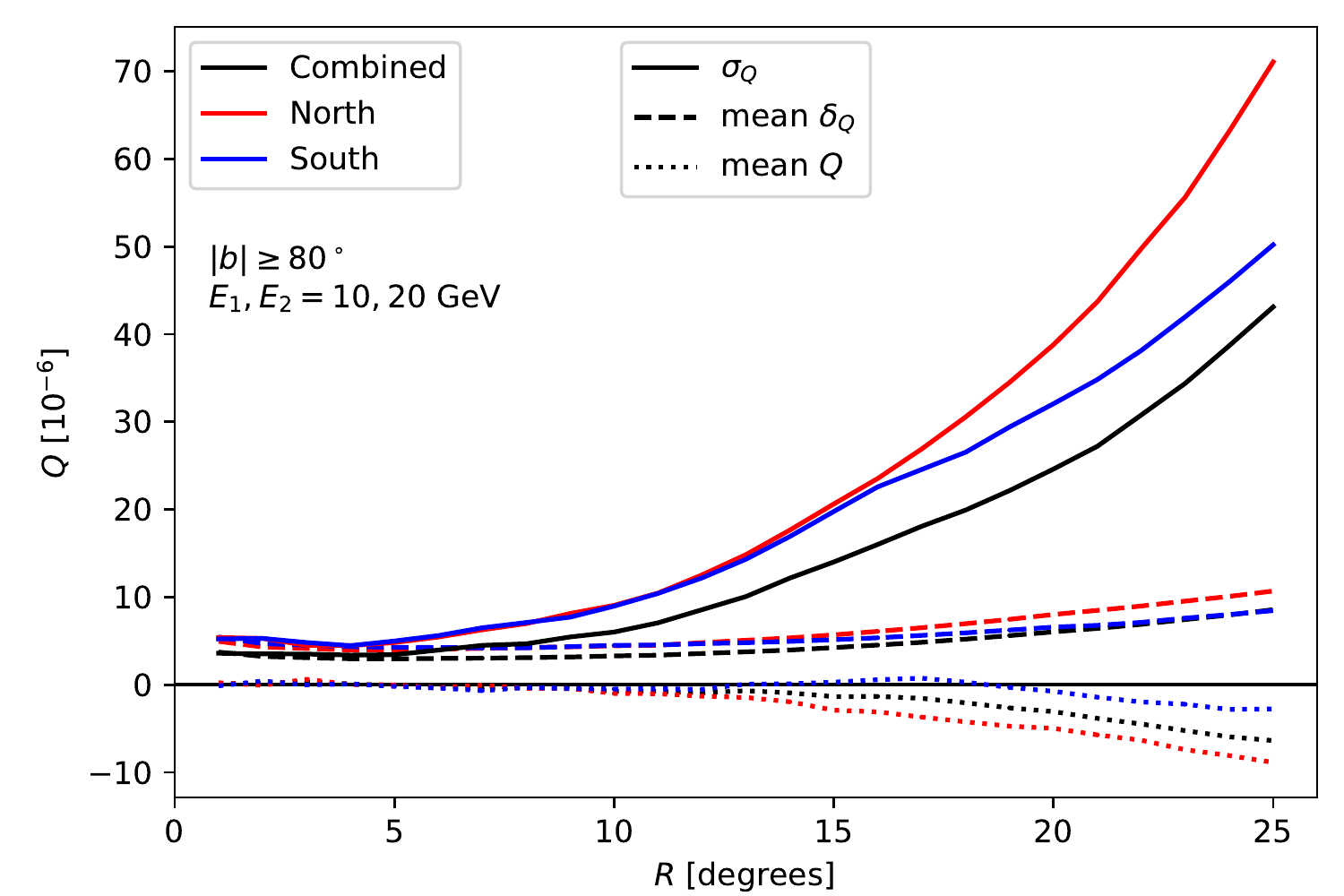}\\
  \includegraphics[width=0.49\textwidth]{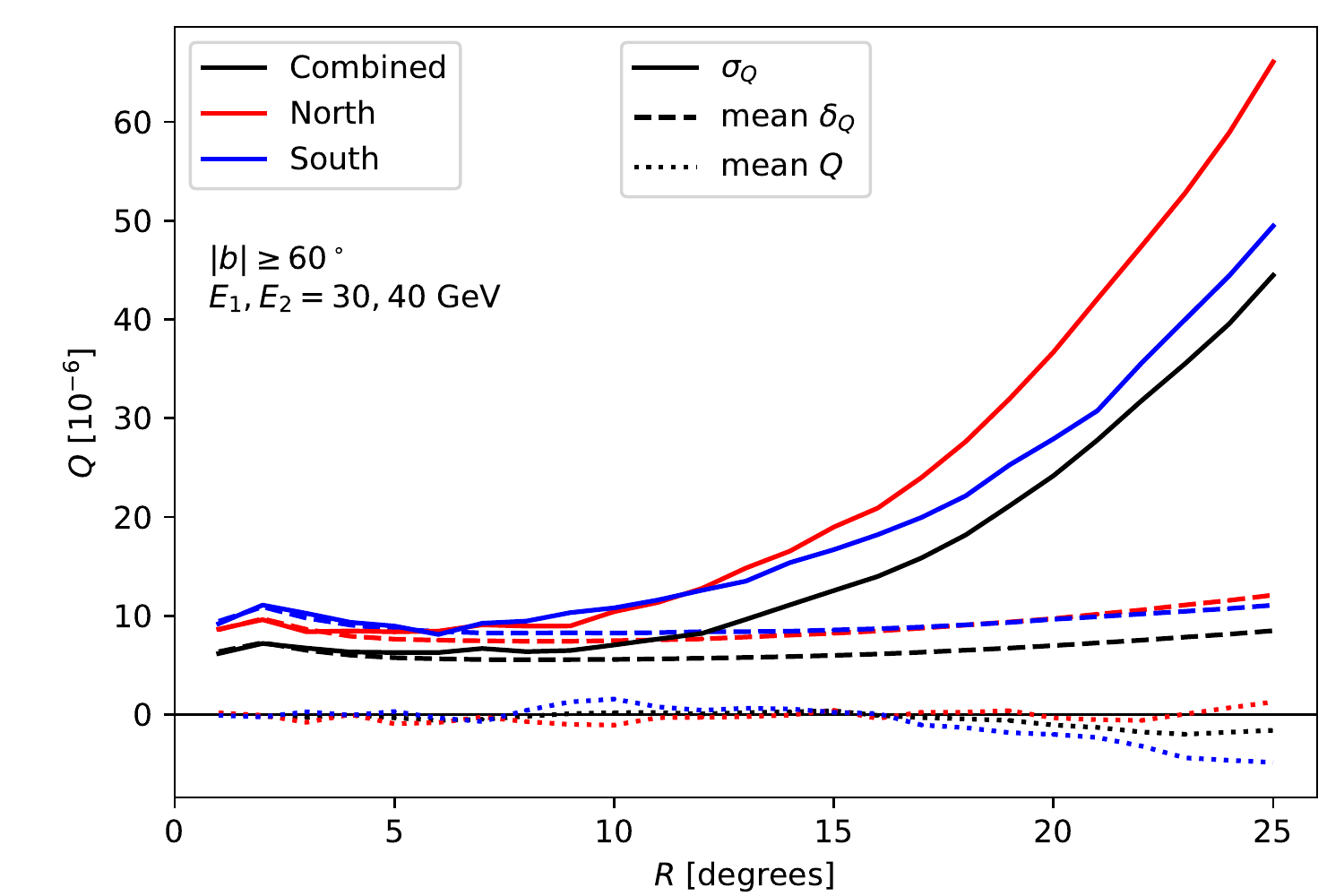}
  \includegraphics[width=0.49\textwidth]{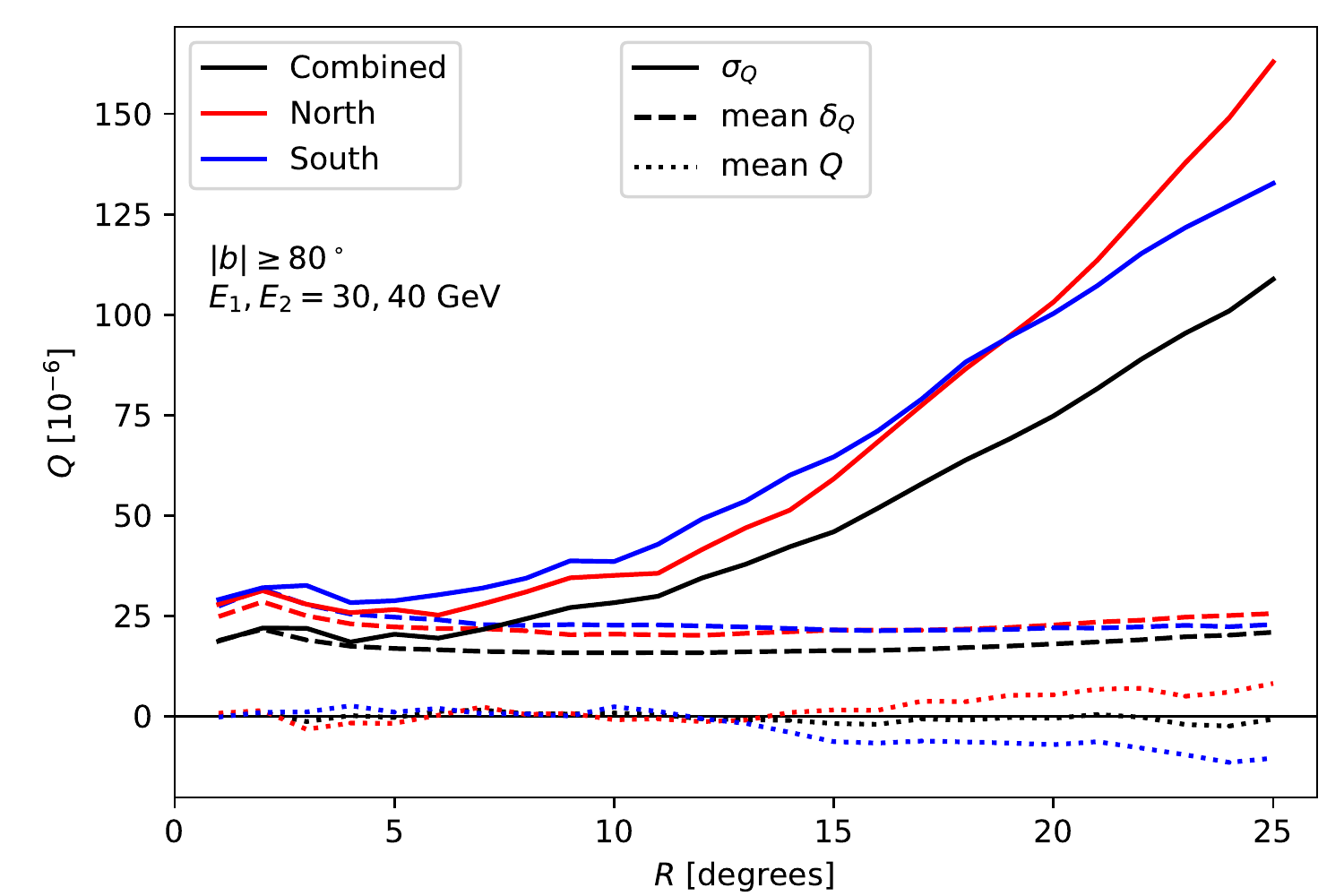}
  \caption{Summary statistics for the 200 simulations of an isotropic sky
  accounting for the LAT instrument response and observing profile. The two
columns represent different latitude cuts and each row a different energy bin
combination.  Shown are for each $R$ the mean value of $Q$ (dotted curves), the mean value of $\delta_Q$
(dashed curves), and the standard deviation $\sigma_Q$ (solid curves) 
of the obtained distribution from the simulation.  
Results for the north (red color) and the south (blue color) hemisphere
 are shown independently, as well as the results from the
 hemispheres combined (black color).}
  \label{fig:isotropicstats}
\end{figure*}

The simulations of uniform photon arrival directions are performed to test the code implementation and also
to test the effect of the latitude cut without introduction of any spatial
dependence in the photon distribution on the sphere.  Photon arrival
directions are sampled uniformly on the sphere by sampling independently in
longitude and cosine of colatitude.  Three sets were sampled and the number of
photons in each sample corresponds roughly to the number of photons selected
above $60^\circ$ latitude for the 10, 20, and 50~GeV ranges.  For
these simulations, the number of photons is always the same in each bin,
independent of the latitude cut applied.  For each latitude cut, 500
simulations are performed, each using a different seed for the pseudo-random
number generator.  For each simulation, the value of $Q$ is calculated for $R$
from $1^\circ$ to $25^\circ$ in steps of $1^\circ$.  For each value of $R$,
the mean of the $Q$ values, the mean of the $\delta_Q$ values, and the
standard deviation of the $Q$ values, $\sigma_Q$, are calculated.  If
everything works as expected, then the value of $\sigma_Q$ 
and the mean of $\delta_Q$ should be identical, and the mean value of $Q$ should be zero.

Figure~\ref{fig:uniformstats} shows the summary statistics for the 500 sets of
simulations employing the three different latitude cuts.  The mean value of $Q$
for these simulations is consistent with 0 for all latitude cuts and the mean
value of $\delta_Q$ agrees with $\sigma_Q$ when the latitude cut is the same
for all photon sets.  The only difference between applying a latitude cut and
not in that case is the magnitude of the error which increases significantly
with $R$ if a latitude cut is applied.  With no latitude cut at all a larger patch
radius will always include more photons, uniformly distributed around 
the arrival direction of the selected $E_3$ photon.
The uncertainty of the $Q$ value thus slightly decreases with increasing $R$.
Applying a latitude cut, however, causes both the spread and the value of $\delta_Q$ to
increase with $R$. This is expected for geometrical reasons; when the same
latitude cut is applied for all energy bins, patches near the boundary will not be circular
but have a sharp cut off. This means the photons will not be uniformly
distributed around $E_3$, but concentrated 
on the higher-latitude side of the direction of the photon
leading to a larger
value for those items in the sum of Equation~(\ref{eq:qstatistic}).
This increases $\delta_Q$ and the absolute values of $Q$, which in turn
increases the standard deviation of the distribution of $Q$ values.

In the case where a latitude cut is
applied only to the $E_3$ photons, the values of $\delta_Q$ and $\sigma_Q$
start to deviate for larger values of $R$.  While $\delta_Q$ follows 
a trend
similar to that when no latitude cut is applied, $\sigma_Q$ more closely
resembles the results with a latitude cut, although with a smaller magnitude.
The onset of the deviation depends
on the latitude cut, starting at smaller $R$ for more strict latitude cuts.
This indicates that it is a boundary effect, but a concrete reason for this
behavior is not understood at the moment.  It is, however, clear that
$\delta_Q$ significantly underestimates the statistical uncertainty of the
measurements when a latitude cut is applied only to $E_3$.  Note though that
the statistical uncertainty in this case is still smaller than when applying
the latitude cut to all energy bins and this method is therefore still preferred.
Hereon, the latitude cut is thus only applied to the photons in the highest
energy bins, 
while for the other bins the photon directions are restricted to being within $D(\nn_k,R)$.

\subsection{Isotropic Emission}

\begin{figure*}
  \centering
  \includegraphics[width=0.49\textwidth]{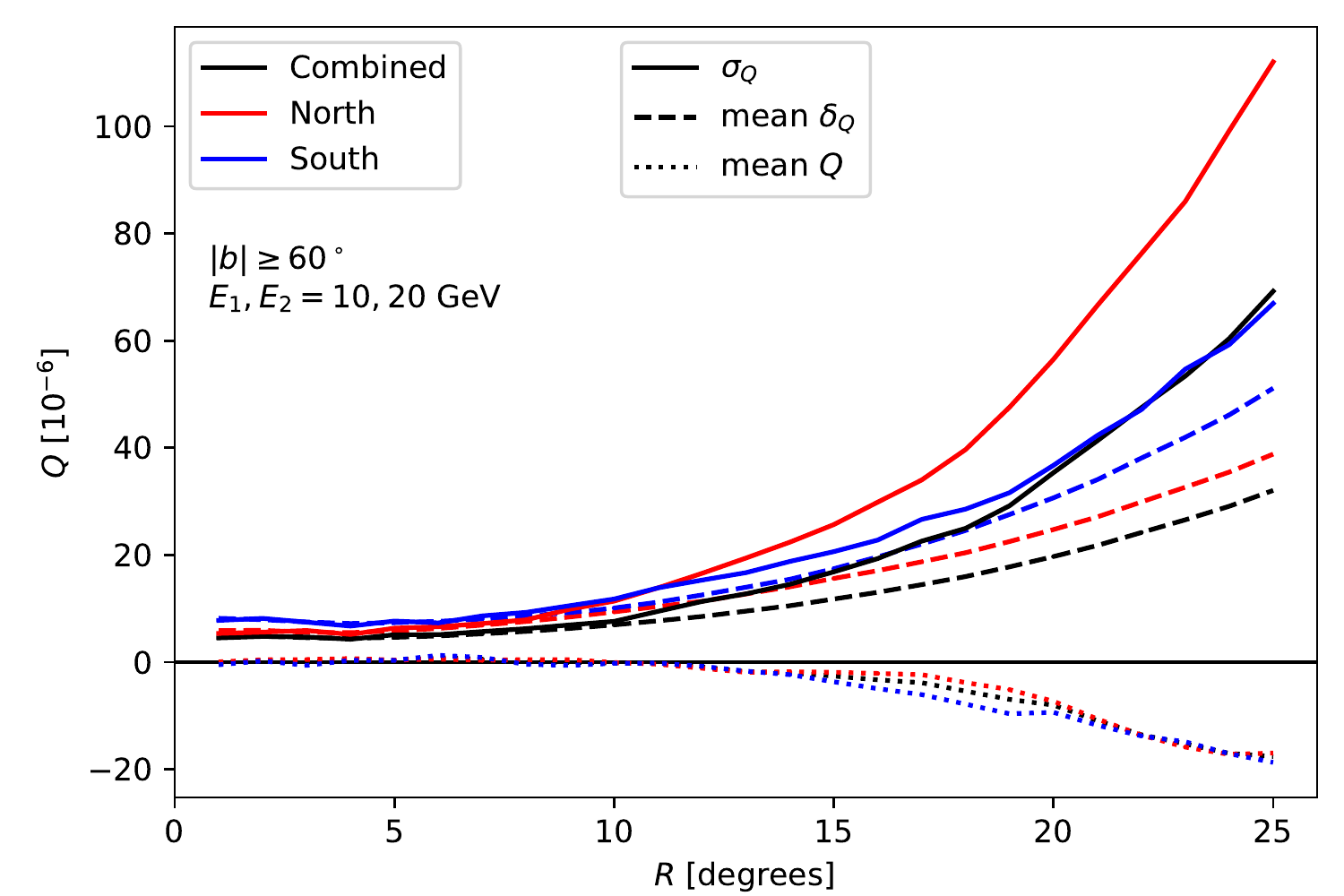}
  \includegraphics[width=0.49\textwidth]{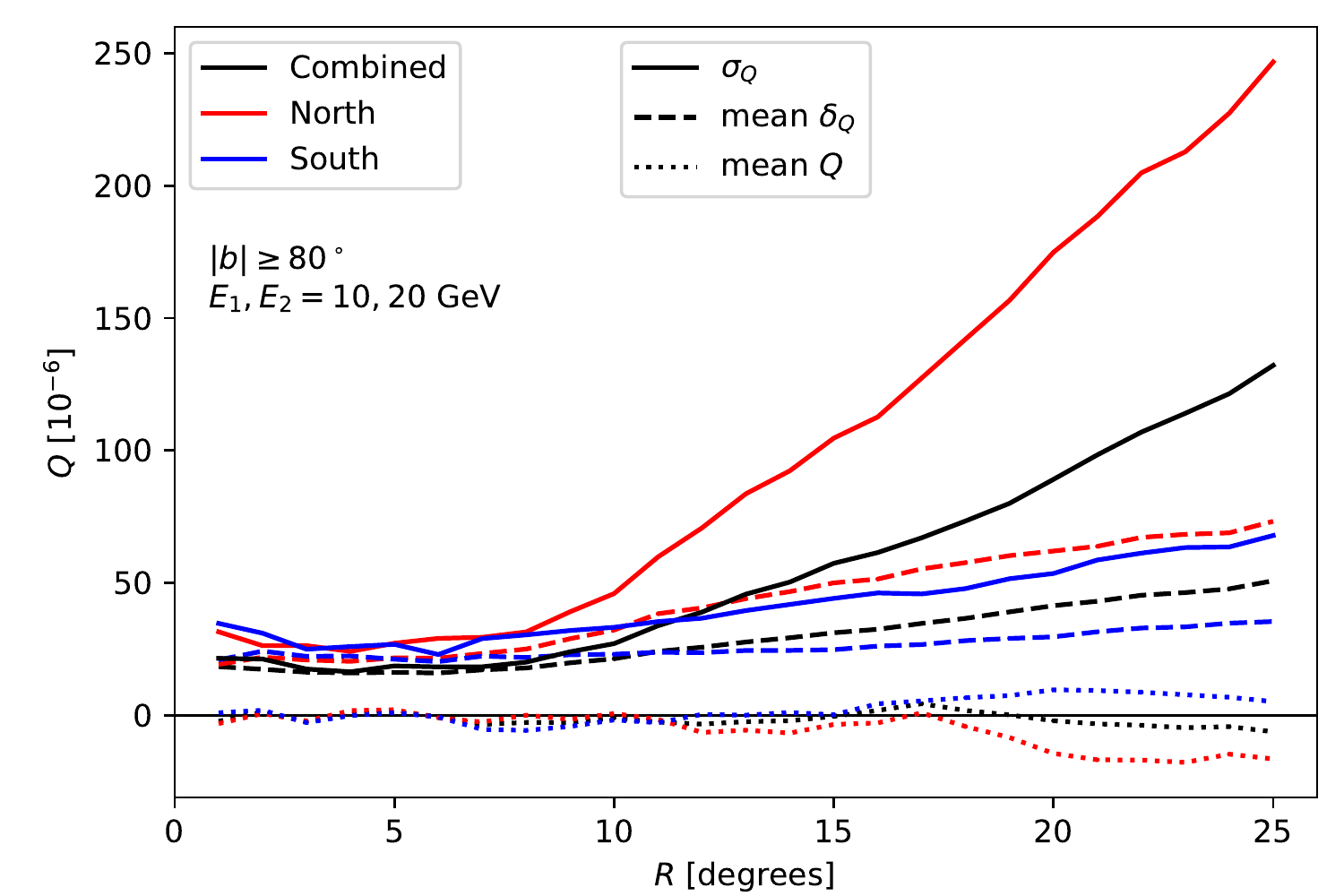}\\
  \includegraphics[width=0.49\textwidth]{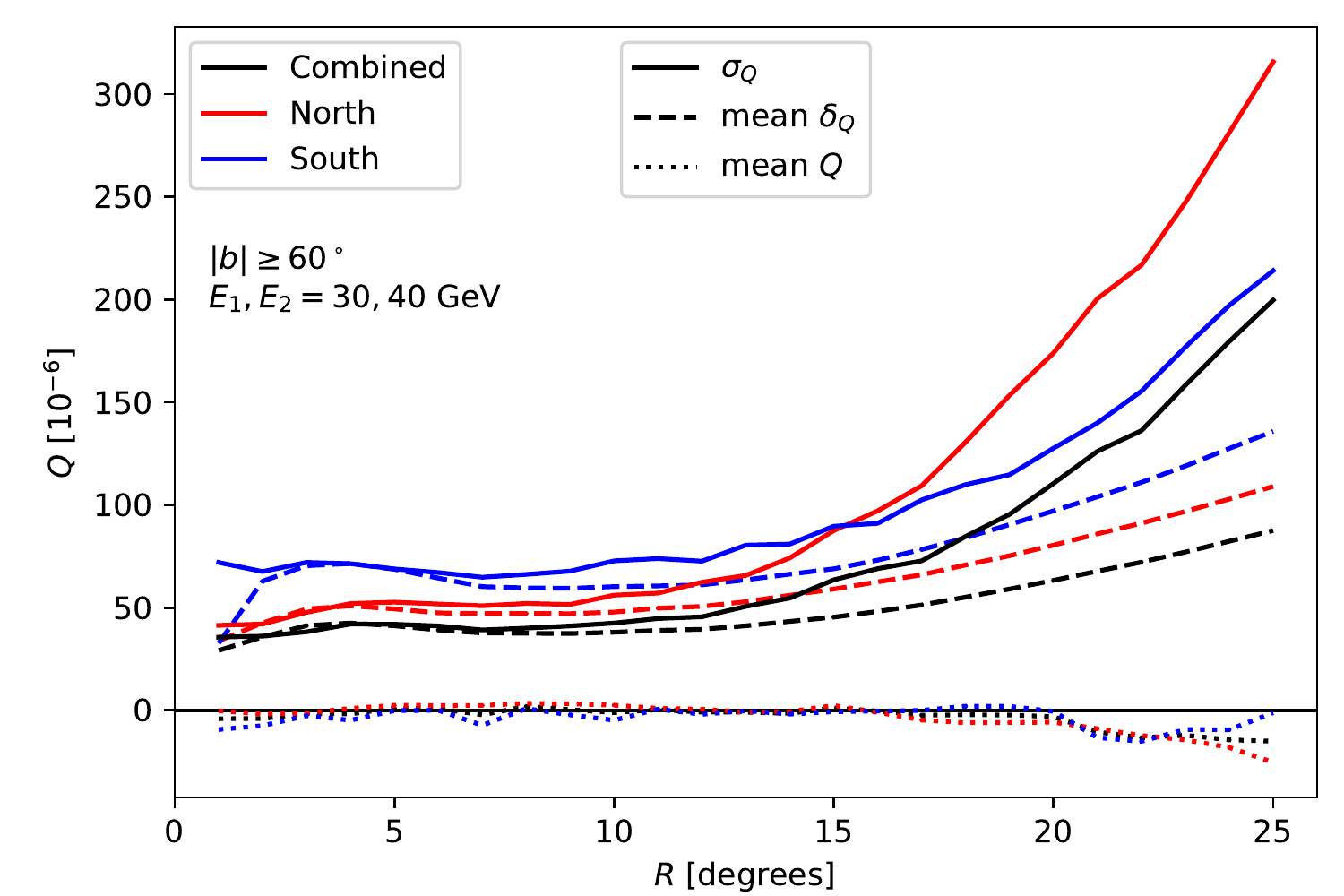}
  \includegraphics[width=0.49\textwidth]{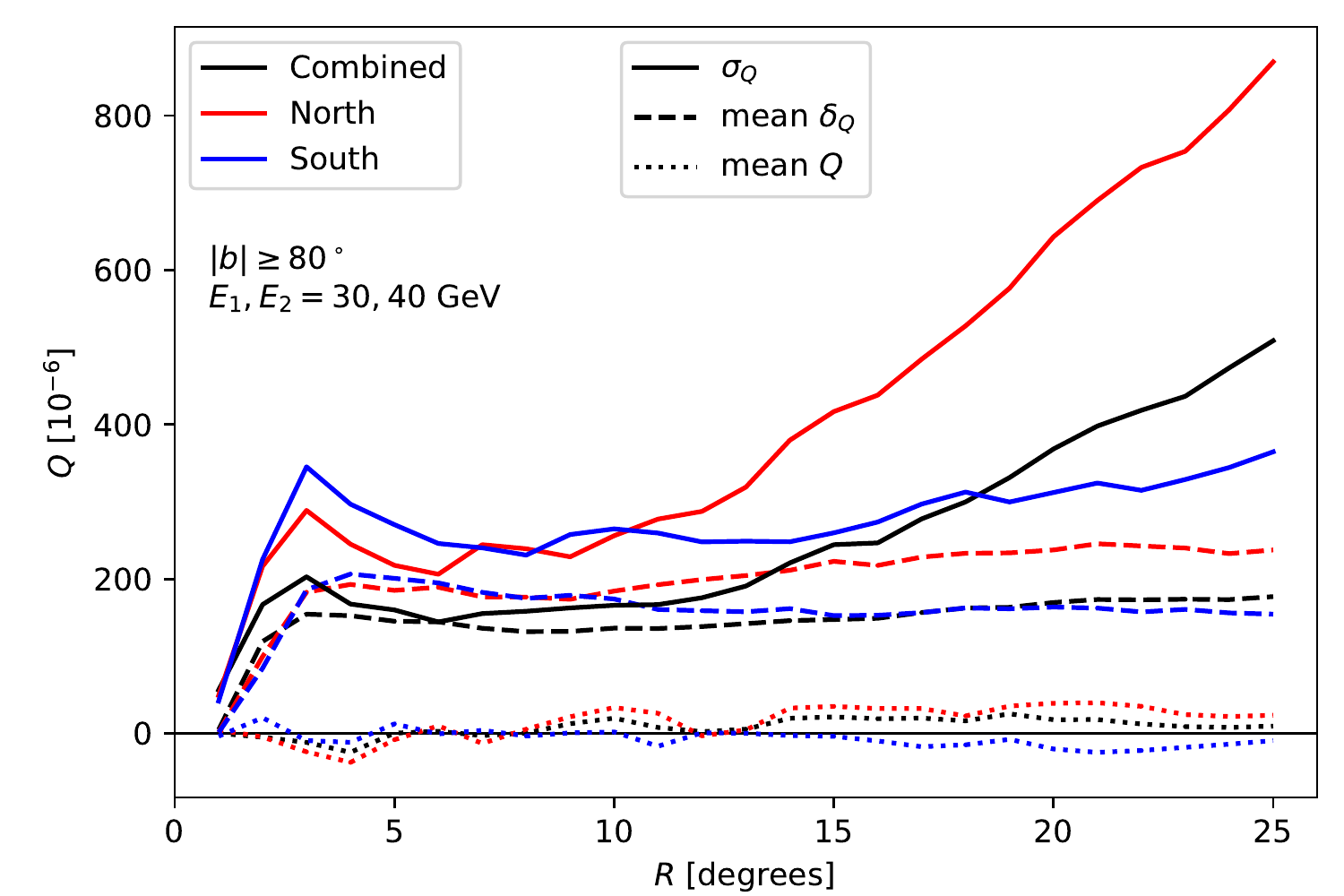}
  \caption{Summary statistics for the 200 simulations of the interstellar
    emission
  accounting for the LAT instrument response and observing profile. The two
columns represent different latitude cuts and each row a different energy bin
combination.  Similar statistics are shown in Figure~\ref{fig:isotropicstats}
for the isotropic emission.}
  \label{fig:iemstats}
\end{figure*}

The unresolved extragalactic emission at GeV energies is approximately
isotropic \citep[e.g.,][]{2015ApJ...799...86A}.  If the 
effective area of the instrument and its exposure were uniform over the sky,
this isotropic emission would lead to a uniform distribution of photon
direction.  This is, however, not the case
and the effect was not studied explicitly by \citet{2014MNRAS.445L..41T}.
The effective area of the
instrument is dependent on both the energy and the incident angle of the
photon which, combined with the observing profile of the LAT, leads to a
non-uniform distribution of photon directions that is slightly energy
dependent.  To test the effect of this on the $Q$ statistics, 200
simulations\footnote{200 simulations are enough to estimate the $1\sigma$
uncertainty with reasonable accuracy.}
were performed using \texttt{gtobssim} 
\footnote{\url{https://fermi.gsfc.nasa.gov/ssc/data/analysis/scitools/overview.html}}
accounting for the true observing
profile of the LAT for the same period as the observed data. The input model
assumed isotropic emission with a power-law distribution in energy 
having an index of $-2.3$ that approximately matches the emission given in `\texttt{iso\_P8R3\_SOURCE\_V2\_v1.txt}'
\footnote{\url{https://fermi.gsfc.nasa.gov/ssc/data/access/lat/BackgroundModels.html}}.
Care was taken to assign a unique seed to each simulation and, because the
simulations use a realistic emission model, the 
numbers of photons are similar to the counts in the LAT data.

Figure~\ref{fig:isotropicstats} shows the results from these simulation using
similar summary statistics as in Figure~\ref{fig:uniformstats}.  The only
difference here is that the results are shown separately for the north and the
south pole to see if there is any hemispherical difference.  In their original
work, \citet{2014MNRAS.445L..41T} found the signal to be more significant in
the northern hemisphere than in the southern one, so it is important to see if the
exposure causes this.  Comparison
between the results of the uniform photon distribution and the isotropic
emission reveals interesting similarities, but also differences.  The latitude
cut applied to the data results in a similar deviations between $\sigma_Q$ and the mean of
$\delta_Q$, but the effect of the exposure leads to an even
larger discrepancy.  Also, rather than being constant with
increasing $R$, the value of $\delta_Q$ rises slightly with $R$.  When
comparing the results for the two hemispheres separately, it is clear that
$\sigma_Q$ is consistently larger in the north compared to the south at larger $R$.
This is despite the northern hemisphere containing more photons in the
simulation than the south.  It is not
clear whether the non-uniformity of the arrival direction causes these
changes, but the conclusion is that simulations are required to get an
accurate estimate of the true uncertainty of the measurement of $Q$.

Because the exposure of the LAT is slightly energy dependent
in the energy range considered, there is the
possibility that it causes a bias in the determination of the value of $Q$.
While there seems to be a small deviation from zero and therefore a small bias in the results shown in
Figure~\ref{fig:isotropicstats}, detailed investigations of the individual
simulations show that these are caused by single outliers 
and the median value is closer to zero.
To distinguish between the effects of limited statistics and a proper bias,
calculations of the $Q$ values from the binned
exposure maps under the assumption of ``infinite'' statistics
were performed by using the pixel locations as photon directions and the pixel values as photon ``counts''.
This resulted in biases
that were orders of magnitude smaller than indicated by the simulations and
the value of $Q$ is therefore not biased by the exposure.

\subsection{Interstellar Emission}
\begin{figure*}
  \centering
  \includegraphics[width=0.49\textwidth]{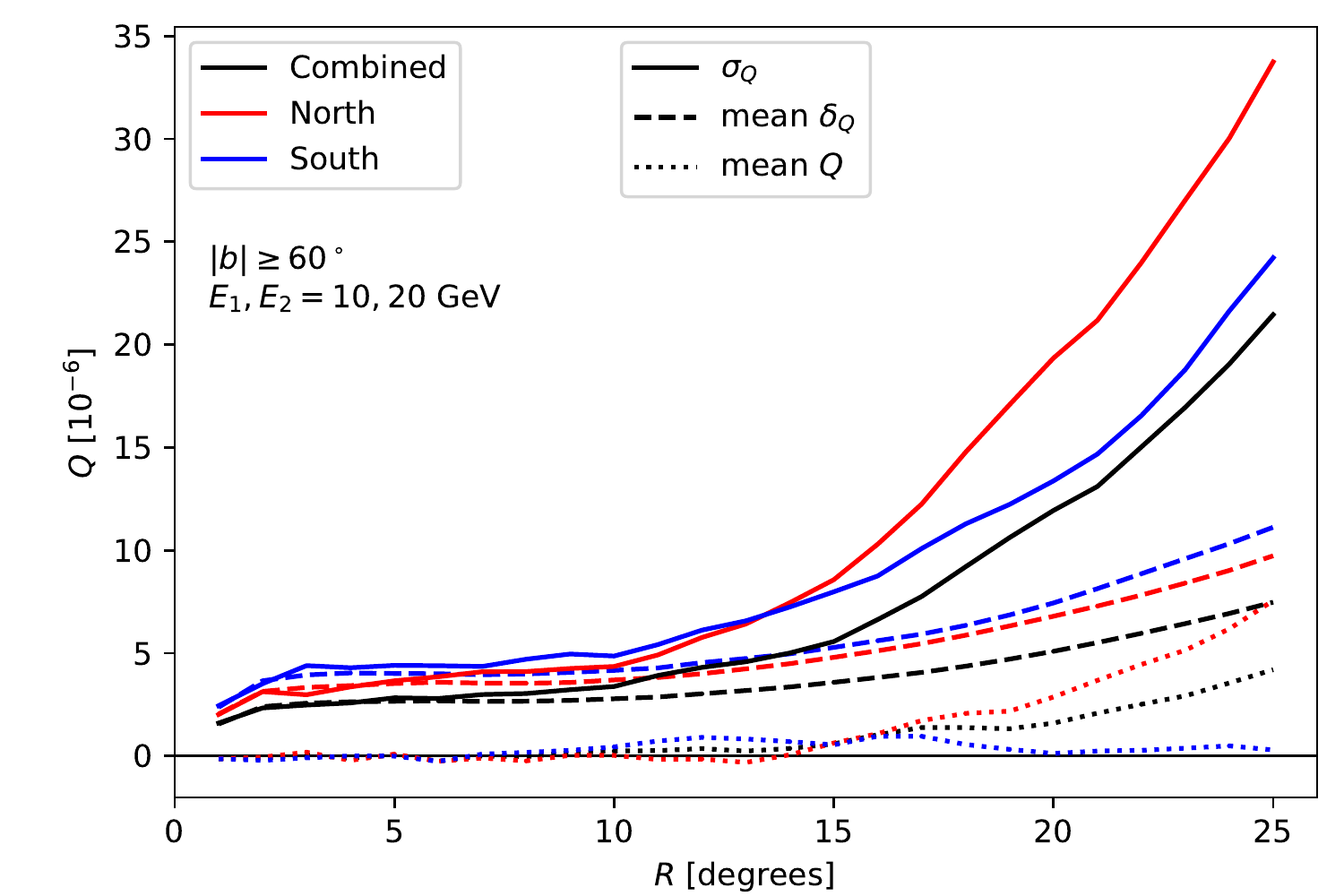}
  \includegraphics[width=0.49\textwidth]{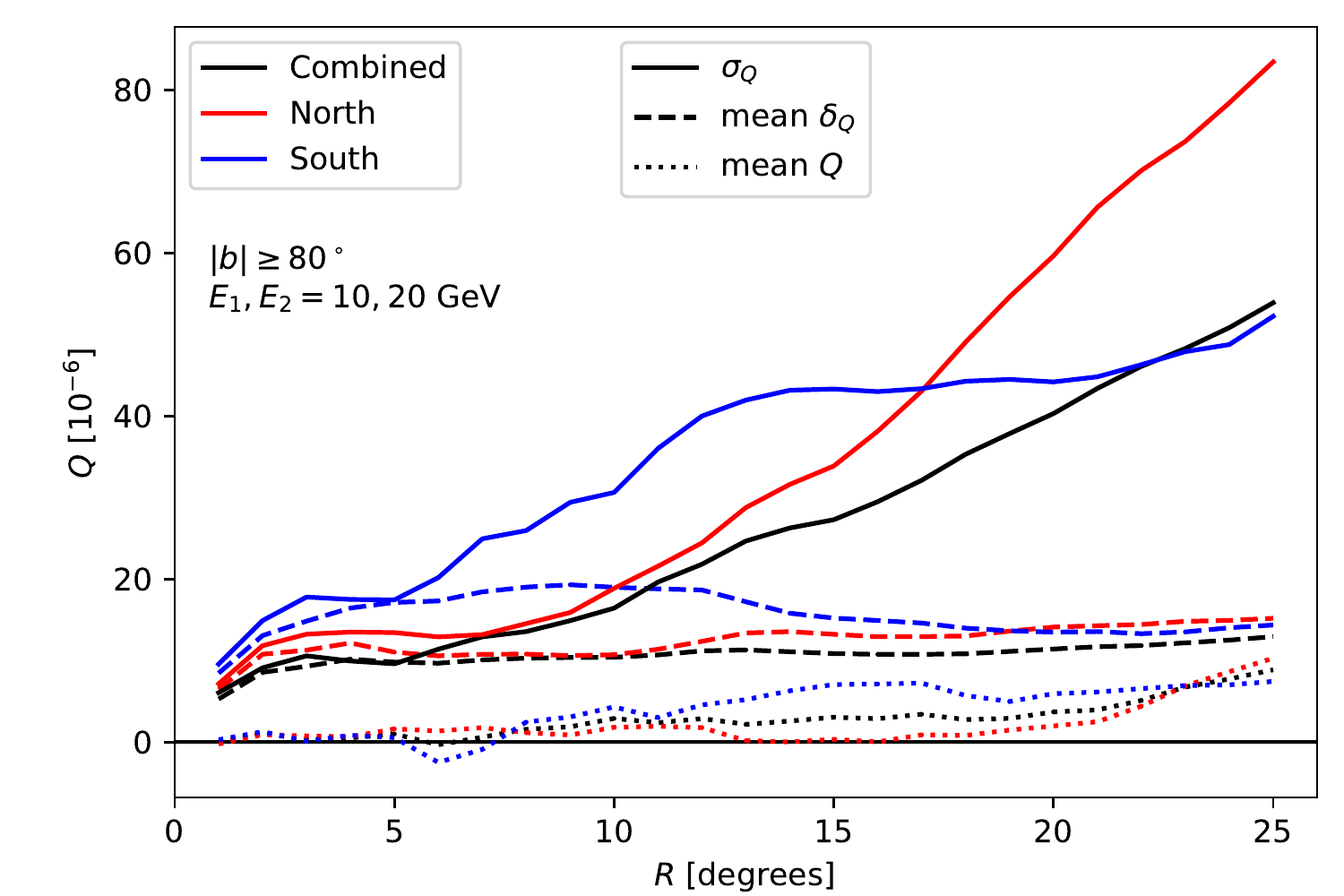}\\
  \includegraphics[width=0.49\textwidth]{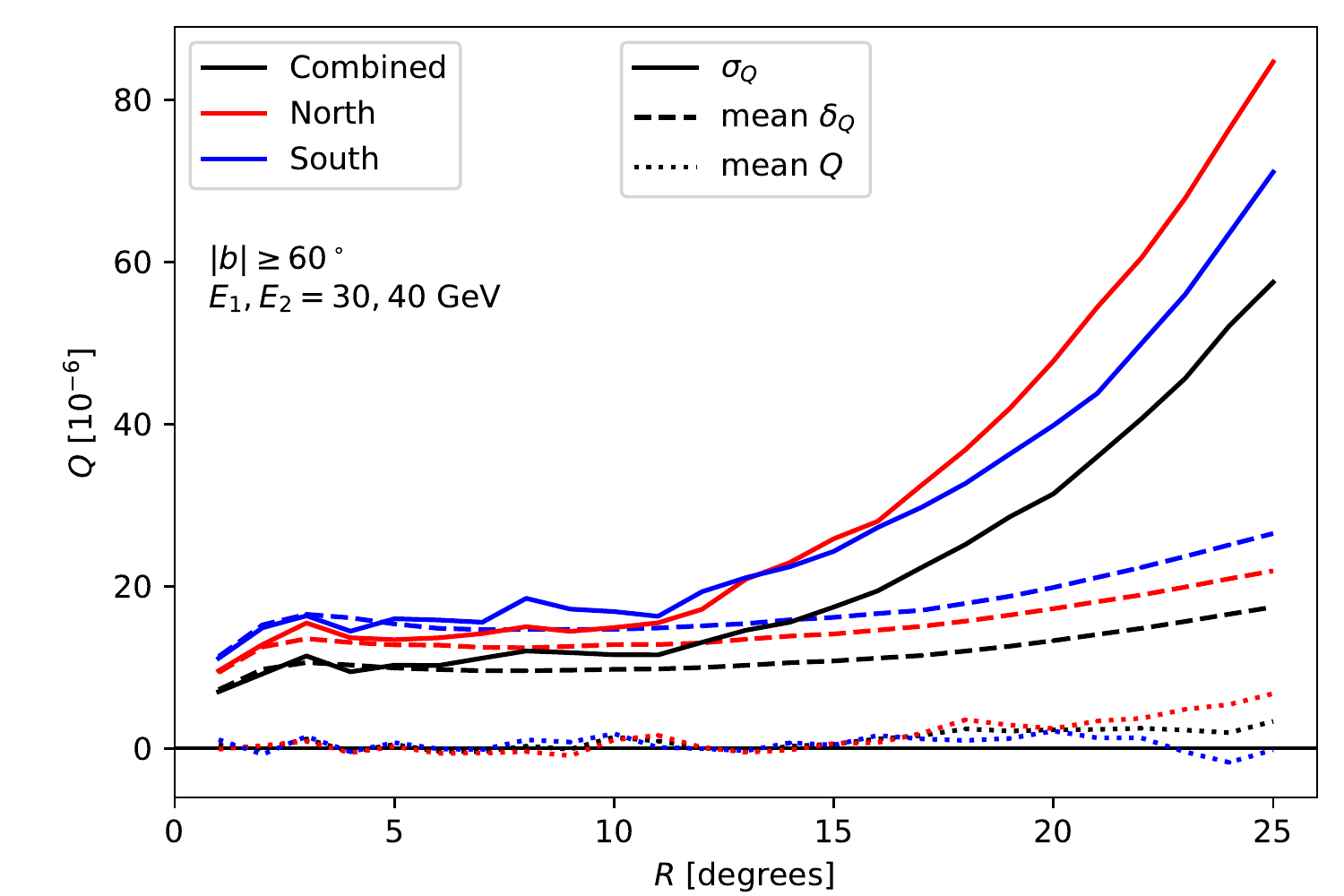}
  \includegraphics[width=0.49\textwidth]{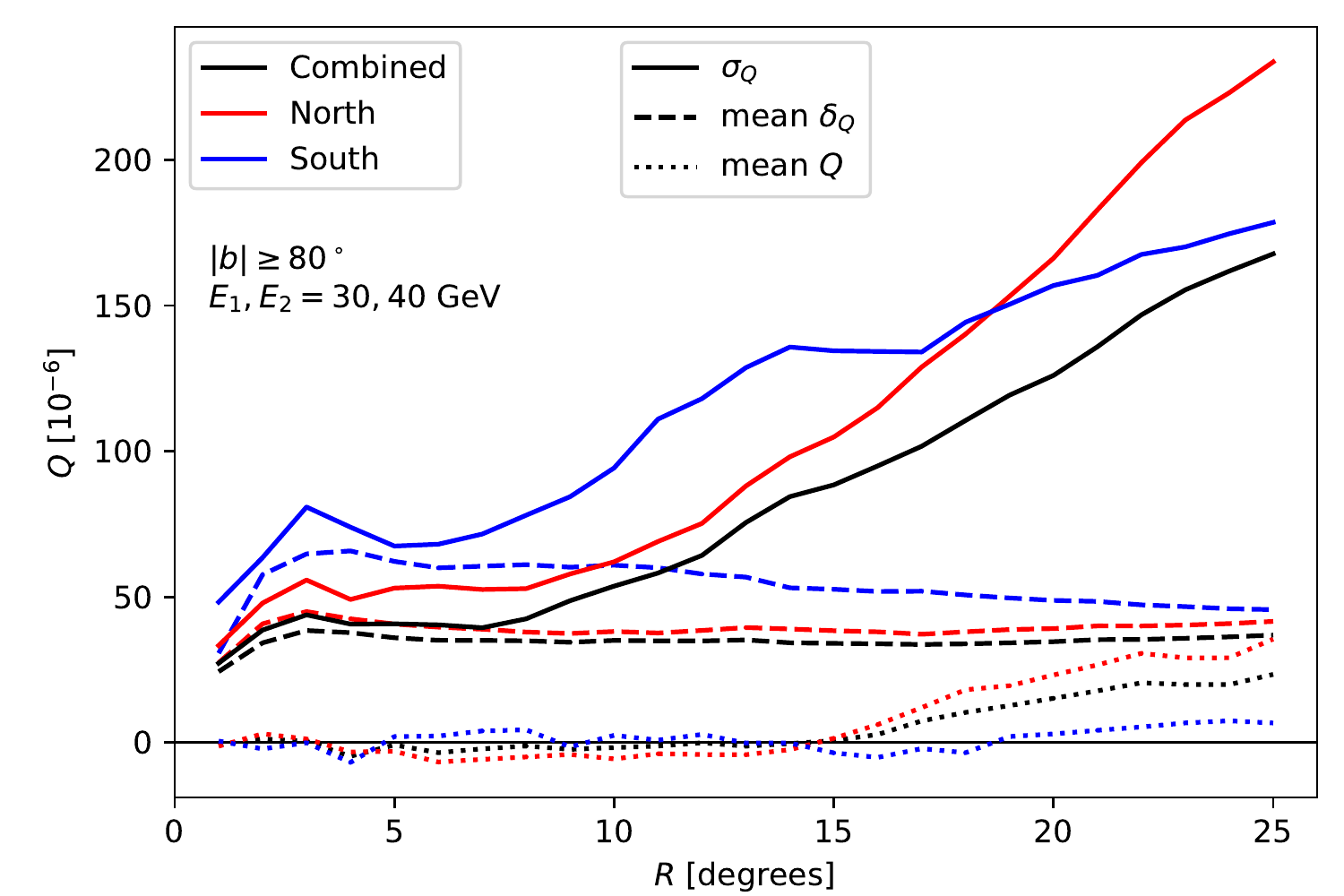}
  \caption{Summary statistics for the 200 simulations of the combined
    interstellar and isotropic emission
  accounting for the LAT instrument response and observing profile. The two
columns represent different latitude cuts and each row a different energy bin
combination.  Similar statistics are shown in Figure~\ref{fig:isotropicstats}
for the isotropic emission only and in Figure~\ref{fig:iemstats} for the
interstellar emission only.}
  \label{fig:combinedstats}
\end{figure*}

Another important consideration is the interstellar emission.
Despite the usage of latitude cuts to reduce its contribution, the
interstellar emission is still a large fraction of the total observed
emission.  Because of its origin in interactions between cosmic rays and the
interstellar medium, the interstellar emission is very structured and some of
that structure is energy dependent.  It has thus the potential to introduce
a bias in the $Q$ statistic.  To test this, 200 simulations were
performed using \texttt{gtobssim}, this time using the interstellar emission
model \texttt{gll\_iem\_v06.fit} 
\footnote{\url{https://fermi.gsfc.nasa.gov/ssc/data/access/lat/BackgroundModels.html}}
as input.

The summary statistics for these simulations are shown in Figure~\ref{fig:iemstats}
for the same latitude cuts and energy bins as used in
Figure~\ref{fig:isotropicstats} for the isotropic simulations.  The most noticeable
difference is the significantly larger range of $\sigma_Q$ compared to the isotropic
simulations.  This is expected, because the interstellar emission is less
intense than the isotropic emission and it also falls off more quickly with
energy and latitude, leading to fewer photons for the evaluation of the $Q$ 
statistic and hence larger statistical errors.  For example in the 10~GeV bin, the
number of photons is similar in the simulations of the two emission models for
the latitude cut of $60^\circ$ while with a cut of $80^\circ$, the number of
photons for the interstellar emission simulation is only half of that using
the isotropic emission.  Also, the number of photons in the 50 GeV bin in
the interstellar emission simulations is always much smaller than that in the
same bin in the isotropic simulations, with only about 10 photons in each
hemisphere when a latitude cut of $80^\circ$ is applied.

Another interesting trend to notice is the 
dependence of the mean of $\delta_Q$ on $R$ for the two different latitude
cuts.  The looser cut of $60^\circ$ results in the values increasing with $R$
while for the $80^\circ$ cut, the values are nearly constant and more in line
with the results from the isotropic simulations shown in
Figure~\ref{fig:isotropicstats}.  It thus seems that the structure in the
interstellar emission increases the uncertainty of the
measurement with increasing $R$ and that this increase is dependent on latitude.
The latitude dependence is of course expected and is the reason for applying a
latitude cut in the first place.  Finally, the hemispheric dependence of the
uncertainties at large $R$ is striking.  In many cases, the southern hemisphere
shows smaller uncertainties than the combined emission, meaning that something
about the structure of the emission is causing a large scatter in the
calculations.  This is despite the fact that the
combined analysis uses about twice as many photons than that
in the southern hemisphere and demonstrates the importance of using these
simulations to estimate the uncertainty of the measurements.

There is a hint of a bias in the determination of $Q$ from the interstellar
emission model.  The mean value of $Q$ from the 200 simulations 
is clearly offset from 0 at higher values of $R$ for most of the permutations of energy
bins and latitude cuts (not all are shown here).  The bias is negative in all
cases where it can be seen
(e.g., the top left panel in Figure~\ref{fig:iemstats}),
which is in contrast to the simulations of the
isotropic emission that showed both positive and negative biases.  To verify 
this, a calculation of $Q$ was performed using
``infinite'' statistics, basically calculating the value of $Q$ based on the pixel
values in the input map \texttt{gll\_iem\_v06.fit}.  Those calculations
confirmed a small bias in the calculation at the level of $\sim 10^{-6}$.
The bias is seen to increase with $R$ and always be negative. It
is thus smaller than the statistical uncertainty and the larger indications of
biases shown in Figure~\ref{fig:iemstats} are a result of statistical
fluctuations in the simulations.  It should also
be emphasized that the bias will be much reduced in the more realistic
simulations that include both the interstellar and the isotropic emissions
discussed in the next subsection, because the isotropic emission provides the larger fraction of photons.

\subsection{Combined Emission}
\label{sec:combined}

As has been shown in the previous subsections, $\delta_Q$ is not a reliable
estimator of the statistical uncertainty of the results.
To create simulated diffuse emission data as realistic as possible,
the simulations of the isotropic and interstellar emission described in the previous
subsections are combined one by one.
For the proper estimation of the
uncertainty of the measurement it is important to have similar numbers
of photons in each simulation and the observed data.  
Even accounting for the effects of the point source mask, the number of
  photons in the simulations is slightly larger than that in the data.  The
  exact ratio of the two depends on the latitude cut and the energy range, and is
  different between the two hemispheres.  To calculate the ratio, the average
  numbers of photons in the 202 simulations were compared to the actual number
  of photons in the data. The ratio varies from being nearly 1
  in the northern hemisphere for the 50~GeV bin and the latitude cut of
  $b>80^\circ$ to being 0.65
  in the southern hemisphere also at 50~GeV and the latitude cut of
  $b<-80^\circ$.  In general, however, the ratio is around 0.95 in the north
  and 0.9 in the south.  This discrepancy is due to the
  models not accurately representing the data.  In particular, the north--south
  asymmetry is well known and cannot be accounted for by the current
  interstellar emission models in combination with an isotropic background
  \citep{2012ApJ...750....3A}.  To account for this difference, the ratios are
  used to determine the fractions of photons that are removed from the
simulations by random selection.  It was found that accounting for this
increased the uncertainty estimate by up to 20\%, the increase being largest
in the south for the tightest latitude cut.
The $\sigma_Q$ results from these combined
simulations are used as the statistical uncertainties of the
calculations for the observed LAT data.

\begin{figure*}
  \centering
  \includegraphics[width=\textwidth]{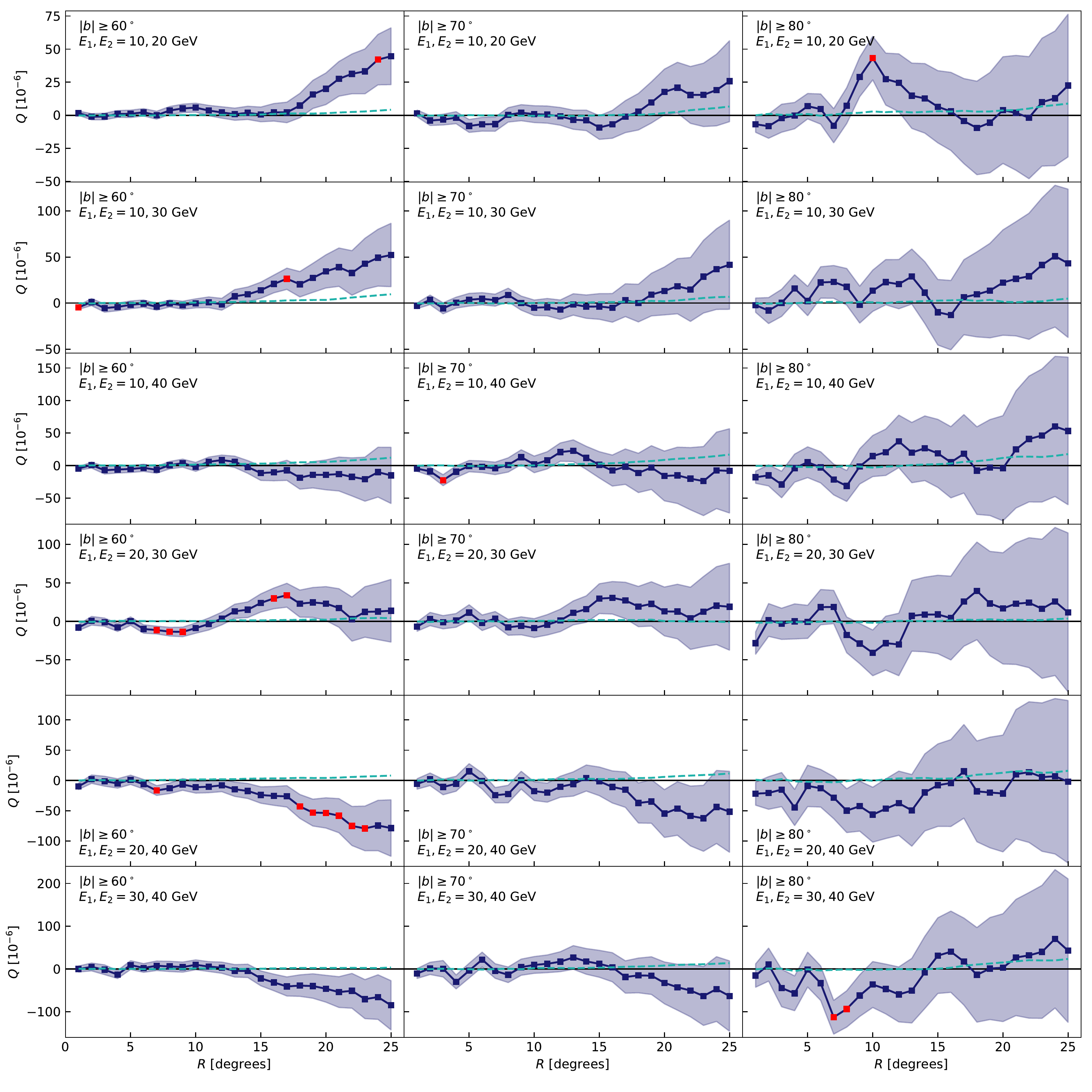}
  \caption{The value of $Q$ calculated using the LAT data is shown as blue
  boxes and the shaded region represents the uncertainty of the measurements
  as estimated from the combined simulations described in
  Section~\ref{sec:combined}. Also plotted is the
mean value of $Q$ from the combined simulation as a dashed cyan curve.
Points that
deviate by more than 2$\sigma_Q$ away from the simulation mean are shown red.  Each column
represent a different latitude cut of $60^\circ$, $70^\circ$, and $80^\circ$
from left to right and each row has a different combination of energy bins as
indicated in each panel.}
  \label{fig:finalResults}
\end{figure*}

Figure~\ref{fig:combinedstats} shows the summary statistics for the combined
simulations.  Not surprisingly, the results are very similar to those for the
isotropic emission only (Figure~\ref{fig:isotropicstats}) because the isotropic
emission is dominant.  There are, however, notable differences mostly
caused by the reduced statistics because of the source mask.  For the
latitude cut of $60^\circ$ and 
combinations using $E_1, E_2 = 10, 20$~GeV, the value of $\sigma_Q$ is 
nearly a factor of 2 larger for $R<5^\circ$, but the difference is smaller
  at larger $R$.  For the $E_1, E_2 = 30, 40$~GeV combination, the fractional change is
  similar at small $R$, but at large $R$ the value of $\sigma_Q$ is about 25\%
larger in the combined simulations.
This reflects the steeper spectrum
of the interstellar emission that is more important at low energies.  For the
tighter latitude cut of $80^\circ$, this effect is not seen and the value of
$\sigma_Q$ is slightly larger in the combined simulation than it is
in the isotropic simulation
due to the source mask.  The $R$ dependence is also different, in
  particular for the south where the value of $\sigma_Q$ is significantly
larger.
Comparison of the values of $\sigma_Q$ and the
mean of $\delta_Q$ shows that the latter starts to underestimate the
statistical uncertainty for values of $R$ between about $5^\circ$ and
$10^\circ$.  The difference is small at first, but rises up to a factor of 2
to 3 at $15^\circ$ and to a factor of 3 to 5 at $20^\circ$.  Without a proper
estimate of the uncertainty, the significance of
results at large $R$ can thus be significantly
overestimated.

The possible bias seen in the results in Figure~\ref{fig:combinedstats} in the
mean of the $Q$ values is a statistical fluctuation in the simulation caused
by strong outliers rather than a real effect.  The bias is also much
smaller than $\sigma_Q$.  Using the uncertainty of the
mean as an estimator for the statistical significance of the bias in the simulations
results in it being less than a 2$\sigma$ effect.  Given that there
are 18 combinations of latitude cuts and energy bins, this could easily be a
statistical fluctuation.  It may, however, indicate that the distribution of the $Q$ values
does not follow the normal distribution and may have more extended tails.
Many more simulations are required to study that in detail.  As will be
shown in the next section, a 1$\sigma$ estimate of the uncertainty is enough at
the moment and further exploration of this is deferred to future work.

\section{Application to LAT data}

\begin{figure*}
  \centering
  \includegraphics[width=0.49\textwidth]{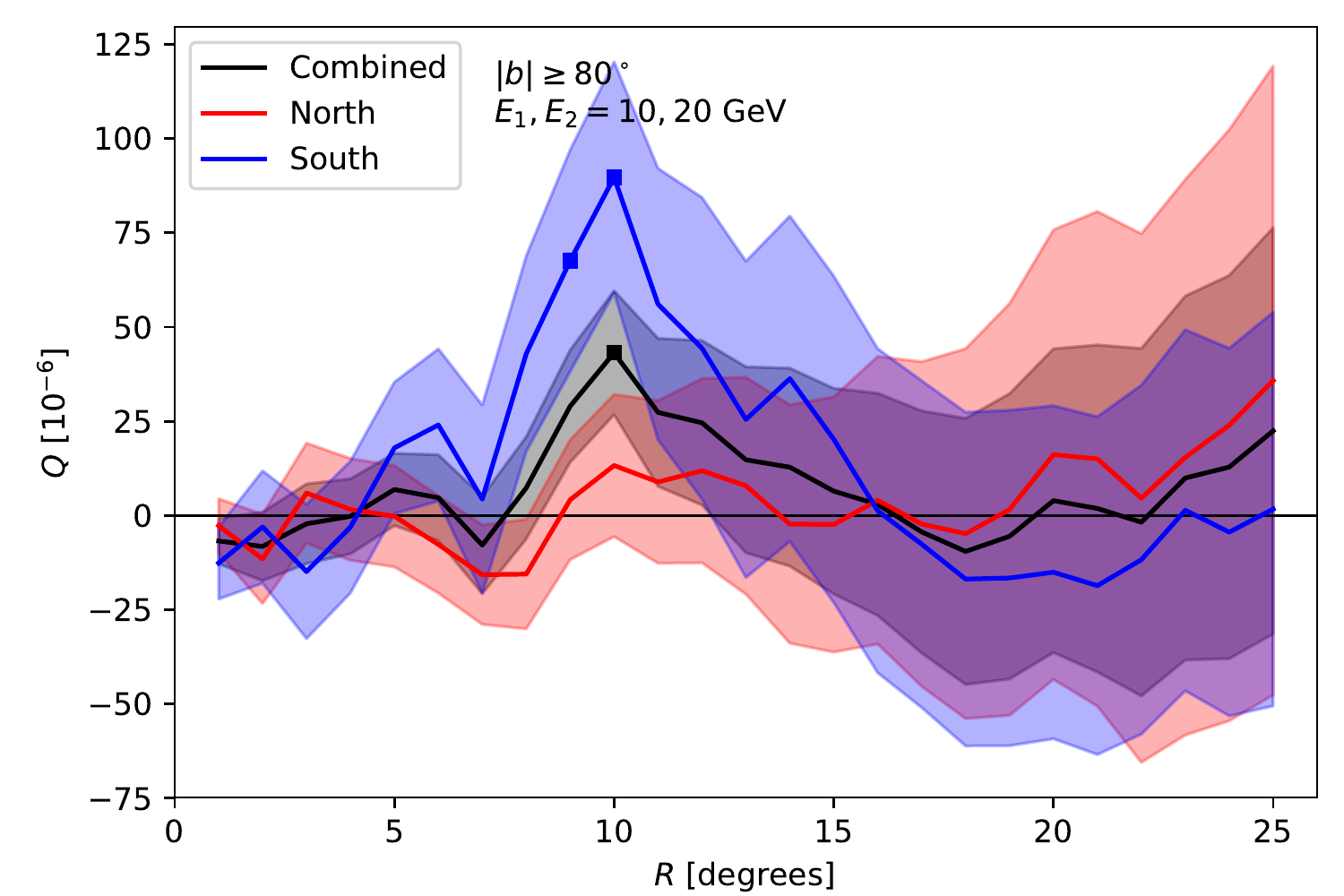}
  \includegraphics[width=0.49\textwidth]{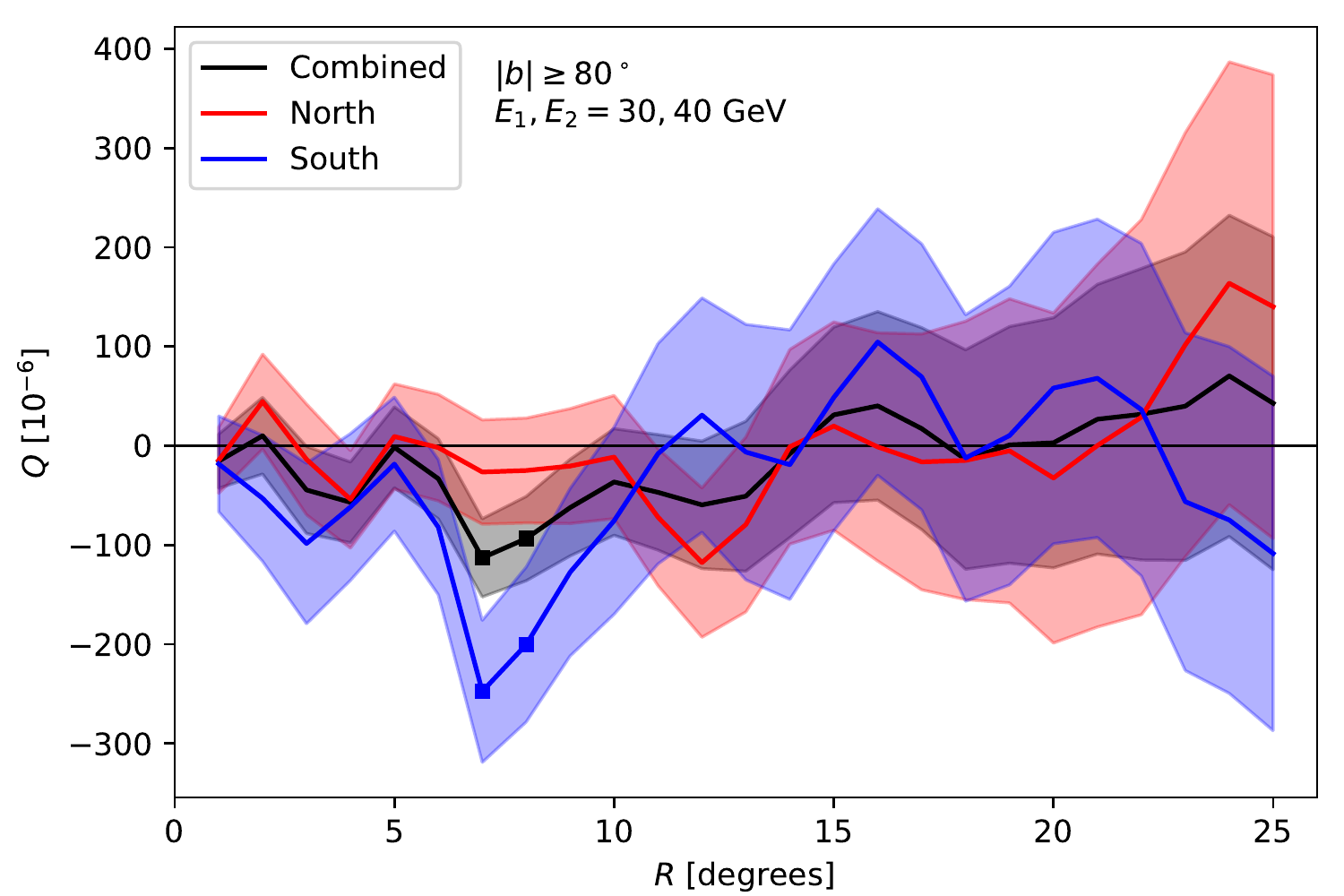}\\
  \includegraphics[width=0.49\textwidth]{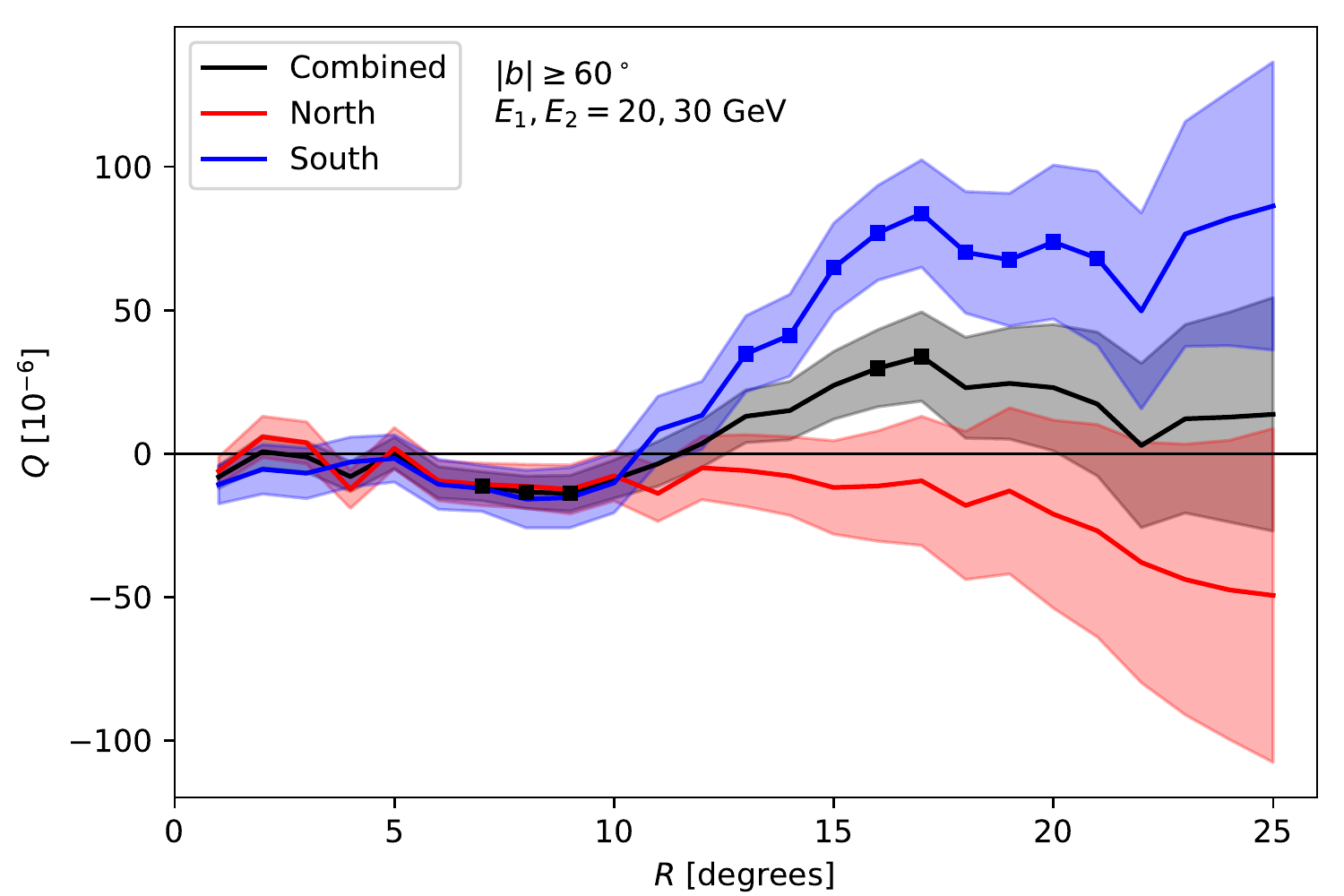}
  \includegraphics[width=0.49\textwidth]{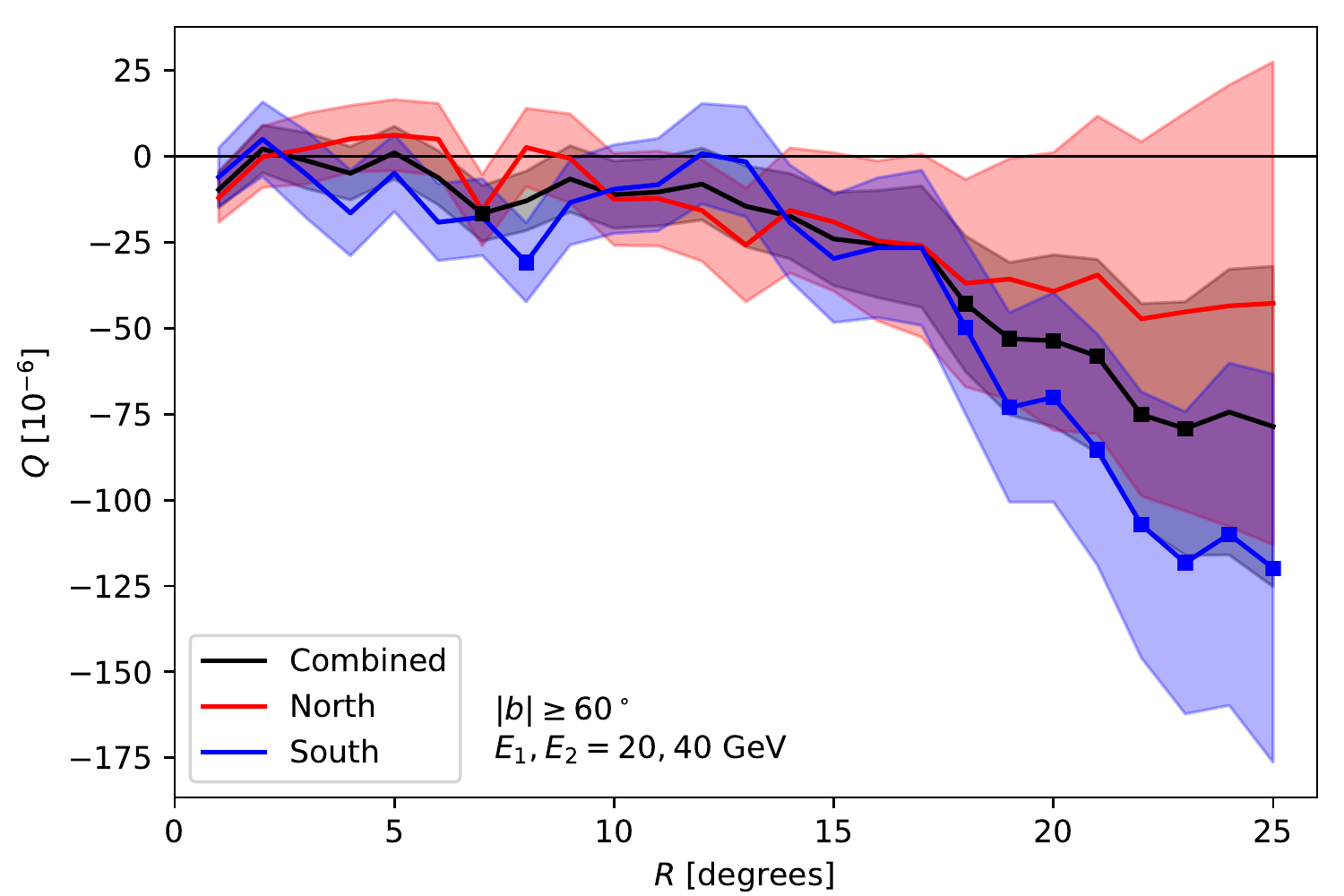}
  \caption{Examples of the results from calculating $Q$ from the LAT data
  separately for the northern and southern hemispheres. The shaded regions
  represent the measurement uncertainties as evaluated from the combined
  simulations described in Section~\ref{sec:combined}.
  The selected latitude cuts and energy bin combinations all have a 2$\sigma$
  outlier in the combined data around $R=7^\circ$, see
  Figure~\ref{fig:finalResults}.  Points that deviate more than
  2$\sigma_Q$ away from the simulation mean are shown as squares.}
  \label{fig:LATresults}
\end{figure*}

The results for the calculation of $Q$ from the observed LAT data 
are given in
Figure~\ref{fig:finalResults}. It shows all combinations of latitude
cuts and energy bins with points that deviate more than 2$\sigma_Q$ from the
simulation mean shown in red.  There is a clear latitude
dependence, with the $60^\circ$ latitude cut showing in many cases an
increasing deviation from 0 with increasing $R$ that is not seen when using
only high-energy events above $80^\circ$.  The sign of the deviation depends
on the combination of energy bins and is likely caused by contamination of
emission from the Galactic plane and it is 
 within 2 $\sigma_Q$ in all but one case where 6 consecutive points are just
  above the limit.  To estimate the significance of this, the fraction of cases
  with 6 consecutive deviations in the simulations was estimated.  This turned
  out to be around 0.9\%, indicating that 0.16 such are expected in our 18
  combinations of latitude cuts and energy bins.  Measuring one when only 0.16
  is expected happens in about 1.2\% of the cases, resulting in a statistical
significance of a little less than 3 sigma.  Apart from this
deviation, only 13 other points are shown in red, of which there is a single
triplet and two pairs.  The simulations were again used to estimate the
fractions of such pairs and the expected number is 0.7 for triplets, 1.6 for
pairs, and 6 for single points.  These 13 points are therefore statistically
consistent with noise.

The large reduction in
significance compared to the results of \citet{2014MNRAS.445L..41T} is caused
by the much improved estimate of the statistical significance through
realistic simulations with \texttt{gtobssim}.  Using only $\delta_Q$ as the
uncertainty would lead to a significant signal in many of the calculations
shown here, but
without a clear trend in handedness.
The most
significant signal seen by \citet{2014MNRAS.445L..41T} was for $E_1,E_2 =
10,40$~GeV and a latitude cut of $80^\circ$, which shows no sign of signal in
the present analysis, even when using $\delta_Q$ as the error estimator.
Their results also indicated that the signal peaked at around $R=12^\circ$,
something that is not seen in Figure~\ref{fig:finalResults}.  In contrast, the
few significant points at $R=7^\circ$ are not visible in Figure~3 of
\citet{2014MNRAS.445L..41T}. Their
results are thus likely caused by statistical fluctuations and were
overstated because of underestimated uncertainties.  
This was already hinted at in the analysis of
  \citet{2015MNRAS.450.3371C}, where the inclusion of the LAT exposure in the
simulations reduced the significance of the signal.

As a check, the data was analyzed separately in the northern and southern
hemispheres.  Splitting the data into twice as many bins does result in more
2$\sigma$ outliers.  
 Most notably though, the number of outliers in the
  northern hemisphere is smaller than expected from the simulations while
  those in the southern hemisphere are more numerous than expected.  This is
  mostly caused by many consecutive points being slightly above the
  2$\sigma_Q$ limit at large values of $R$ for the
  latitude cuts of $-60^\circ$ and $-70^\circ$.  For the former, 5 out of 6
  have these deviations while 3 out of 6 show deviation for the latter.
  No such deviations are seen at the tightest latitude cut of $-80^\circ$.
  The deviations are both positive and negative and in some cases the northern
  hemisphere shows indication of a small signal with the opposite sign to that
  of the southern one.
  The outliers also show a steady rise with $R$ and do not portray any visible
  structure.
  Having so many outliers is statistically unlikely and more simulations are
needed to accurately estimate the statistical significance.  This putative
signal is though unlikely to originate from extragalactic processes, because
it is only seen for one of the hemispheres and only for loose cuts in Galactic
latitude.
A small selection of the results is shown in
Figure~\ref{fig:LATresults} focusing on the few bins with significant outliers
in the combined data.  Examination of the hemispheric dependence further
illustrates that the few outliers visible in the combined data are likely
statistical outliers.  In two cases, the signal is caused by a spike in the
southern hemisphere that is not present in the north while in the other two
it is a fluctuation in both hemisphere.  Examination of the
hemispheric dependence reveals that there is very little correlation between
the two hemispheres, and the largest deviation from 0 in the combined signal
occurs when the signals happen to deviate in the same direction.  There is,
however, no clear trend in the structure of the signal either when looking at
different latitude cuts or different energy bins.  The marginally significant
results are thus likely
caused by statistical fluctuations.

A possible explanation for the missing signal is the usage of the P8R3 SOURCE
data compared to the Pass 7 CLEAN data used in
\citet{2014MNRAS.445L..41T}.  While it is difficult to compare the classes
directly, the P8R3 Source data is expected to have a slightly higher
background rate than the Pass 7 Clean data.  To test the effect of this, the
calculations of the $Q$ values were repeated with the much cleaner P8R3
ULTRACLEANVETO data.  The results are qualitatively similar to those already
presented, but due to the lower acceptance, the statistical power of
the analysis is reduced.  The results are therefore compatible with being statistical
fluctuations and the lack of signal is thus not caused by background
contamination in the SOURCE event class.

Another difference between the current analysis and those of
  \citet{2014MNRAS.445L..41T} and \citet{2015MNRAS.450.3371C} is the use of
  the 4FGL catalog cutting out a $2^\circ$ diameter around the sources instead
  of the LAT first high-energy source (1FHL) catalog
  \citep{2013ApJS..209...34A} with a cut of $3^\circ$ diameter.  Using the
  1FHL catalog would be inappropriate with the larger dataset, but to test the
  effect of this, the analysis was repeated using the LAT third high-energy
  source (3FHL) catalog \citep{2017ApJS..232...18A} and the larger cut.  The
  3FHL is the most recent in the series of high-energy catalogs, using photons with energies between
  10~GeV and 2~TeV for source detection.  It is therefore appropriate for
  analysis in this energy range.  The results are qualitatively consistent
  with the current results and the point source cut does not affect the main
conclusions of this work.

\section{Conclusions}
\label{Concl}
The work of \citet{2014MNRAS.445L..41T} looking for handedness in the
arrival directions of {\em Fermi}-LAT photon data has been repeated with
improved LAT event reconstruction and more data.  Several Monte Carlo simulations were
performed to accurately estimate the uncertainty of the results.
The new error estimate, $\sigma_Q$, is often significantly larger than
$\delta_Q$, the error estimate used in
\citet{2014MNRAS.445L..41T}, resulting in no clear signal of handedness.
As demonstrated here, $\sigma_Q$ better reflects the true spread in the $Q$ values.
It also reveals unexpected boundary effects due to the latitude cuts, which need further investigation.

There is a hint of a nearly 3$\sigma$ signal at large radii for
$E_1,E_2=20,40$~GeV and a latitude cut of $60^\circ$, but the signal is
absent for the tighter latitude cuts and is therefore likely caused by
contamination from Galactic emission.
A similar feature is visible in the results of
\citet{2014MNRAS.445L..41T} for the same energy bin, but the
signal also decreases significantly with more stringent cuts on latitude.
The most promising data selection in \citet{2014MNRAS.445L..41T} using
$E_1,E_2 = 10, 40$ GeV and a latitude cut of $80^\circ$, showed in their
analysis a signal of left handedness with an estimated significance of
about $3\sigma$ but is, in the current analysis, compatible with 0 and
not even a slight hint of a signal at $R=12^\circ$.

This conclusion does not rule out the existence of a helical cosmological
magnetic field. Several assumptions are made in the physical motivation 
presented by
\cite{2014MNRAS.445L..41T}. For instance, there is no way of knowing how many
of the photon triples used actually do originate from the same source. It may
very well be so that there are so few that any signal they might carry is
completely drowned by the background. In other words, the constructed
$Q$ statistic in Equation~(\ref{eq:qstatistic}) may in practice not be as closely
related to the helical part of the correlator of the magnetic field as
assumed.  In fact, \citet{2017JCAP...05..005D} show that random
fluctuations in the magnetic field can induce spurious signals in the
$Q$ statistic and averaging over many realizations is needed to
accurately trace the observed signal back to the helicity of the
magnetic field; even the sign may be incorrectly estimated.  They
propose a modification to the $Q$ statistic that can improve the power
to determine the handedness, but it is unclear if the improvements can
overcome the effects of the unknown structure of the magnetic field.

We emphasize that our main objective was to see whether---independently of
any model or physical assumptions---there exists any handedness in the LAT data.
In view of our new findings concerning the relatively large error bars, our
answer to this question is no.
This does not necessarily imply that any intergalactic magnetic field must be
weak or that the method of \cite{2013PhRvD..87l3527T} is not sensitive enough.
It is possible that specific selection methods in time or shape of the 
photon triplets
could yield a significant result for $Q$.
As discussed in this paper, it is possible that a finite value of $Q$ could be
caused by regions of different sizes and different energy ranges
in which photons accumulate to a density that is higher than the average.
This could either be caused by instrumental effects (for example by a nonuniformity of
exposure) or it could be caused by a handedness of processes within our Galaxy.
A possible candidate could be the Galactic magnetic field.
In such a case, the causal connection with $Q$ would be different from what
was anticipated by \cite{2013PhRvD..87l3527T}.
However, given that there is currently very little evidence for any handedness,
neither globally or locally for 
the northern or southern Galactic hemispheres
these possibilities remain just speculation.

\acknowledgments
We thank Alexander Eid and Omkar Ramachandran for their help at earlier
stages of this project.
This research was supported in part by the Astronomy and Astrophysics
Grants Program of the National Science Foundation (grant 1615100),
and the Swedish Research Council (grant 2019-04234).

The \textit{Fermi} LAT Collaboration acknowledges generous ongoing support
from a number of agencies and institutes that have supported both the
development and the operation of the LAT as well as scientific data analysis.
These include the National Aeronautics and Space Administration and the
Department of Energy in the United States, the Commissariat \`a l'Energie Atomique
and the Centre National de la Recherche Scientifique / Institut National de Physique
Nucl\'eaire et de Physique des Particules in France, the Agenzia Spaziale Italiana
and the Istituto Nazionale di Fisica Nucleare in Italy, the Ministry of Education,
Culture, Sports, Science and Technology (MEXT), High Energy Accelerator Research
Organization (KEK) and Japan Aerospace Exploration Agency (JAXA) in Japan, and
the K.~A.~Wallenberg Foundation, the Swedish Research Council and the
Swedish National Space Board in Sweden.

Additional support for science analysis during the operations phase is gratefully
acknowledged from the Istituto Nazionale di Astrofisica in Italy and the Centre
National d'\'Etudes Spatiales in France. This work performed in part under DOE
Contract DE-AC02-76SF00515.

\bibliographystyle{hapj}
\bibliography{paper}

\end{document}